# A Novel Explicit Filter for the Approximate Deconvolution in Large-Eddy Simulation on General Unstructured Grids: A posteriori tests on highly stretched grids

Mohammad Bagher Molaei[1], Ehsan Amani[1,*], and Morteza Ghorbani[2]

[1]*Department of Mechanical Engineering, Amirkabir University of Technology (Tehran Polytechnic), Iran*

[2]*Sabanci University Nanotechnology Research and Application Center, 34956 Tuzla, Istanbul, Turkey*

**Abstract**

Explicit filters play a pivotal role in the scale separation and numerical stability of advanced Large Eddy Simulation (LES) closures, such as dynamic eddy-viscosity or Approximate Deconvolution (AD) methods. In the present study, it is demonstrated that the performance of commonly used explicit filters applicable to general unstructured grids highly depends on the grid configuration, specifically the cell aspect ratio, which can result in poor filter spectral properties, ultimately leading to large errors and even solution divergence. This study introduces a novel, efficient explicit filter for general unstructured grids, addressing this shortcoming through a combination of a face-averaging technique and recursive filtering. The filter parameters are then determined through a constrained multi-objective optimization, ensuring desirable spectral properties, including high-wavenumber attenuation, filter-width precision, filter stability and positivity, and minimized dispersion and commutation errors. The AD-LES of turbulent channel flow benchmarks using the new filter demonstrate a noticeable improvement in turbulent flow predictions on highly stretched boundary-layer-type grids, particularly in reducing the log-layer mean velocity profile mismatch, compared to simulations using conventional filters. The analyses show that this enhancement is mainly attributed to the sufficient level of attenuation near the Nyquist wavenumber achieved by the new filter in all spatial directions across various grid configurations, among others. The new filter was also successfully tested on unstructured prism grids for the 3D Taylor-Green vortex benchmark.

**Keywords:** Large-Eddy Simulation (LES); Approximate Deconvolution (AD); Explicit Filter; Unstructured grid

---

[*] Corresponding author. Address: Mechanical Engineering Dept., Amirkabir University of Technology (Tehran Polytechnic), 424 Hafez Avenue, Tehran, P.O.Box: 15875-4413, Iran. Tel: +98 21 64543404. Email: eamani@aut.ac.ir

## 1. Introduction

Large Eddy Simulation (LES) has emerged as a pivotal tool for analyzing turbulent flows. It is based on the decomposition of flow properties into two contributions: a large-scale or resolved component, and a small-scale or Sub-Grid-Scale (SGS) one (Pope 2000). This separation is accomplished through the application of a low-pass spatial filter. In the conventional LES framework, the implicit or grid filtering process is intrinsically embedded within the numerical discretization of the differential equations. However, a significant limitation of this implicit filtering approach lies in its inherent inconsistencies, particularly its inability to achieve a well-defined 3D filtering effect (Lund 2003). Moreover, the absence of control over the frequency content of the solution can increase aliasing errors, leading to numerical instabilities, and introduce sensitivity to grid resolution (Vasilyev *et al.* 1998, Chow & Moin 2003, Kim *et al.* 2021). Additionally, the numerical discretization errors associated with spatial derivatives play a significant role in the overall LES error. An analysis of modified wavenumber plots for discretization schemes (Stolz et al. 2002) reveals that these errors are particularly pronounced for wavenumbers approaching the Nyquist frequency. Consequently, minimizing or eliminating the influence of such wavenumbers on the solution can enhance the accuracy and reliability of LES results.

A possible approach to mitigate the grid dependency and discretization errors of the scales close to the Nyquist is the application of an explicit low-pass filter with a cut-off wavenumber substantially lower than the Nyquist wavenumber (Lund & Kaltenbach 1995, Najjar & Tafti 1996, Bose 2012, Kim *et al.* 2021, Amani *et al.* 2024). By applying an explicit filter, an additional scale separation is achieved, creating a Sub-Filter Scale (SFS) alongside the conventional SGS. While the SGS remains inherently unresolved and must be approximated through modeling, the SFS can, in principle, be reconstructed theoretically (Gullbrand & Chow 2003). On the other hand, the low-pass filter plays a pivotal role as the test filter in dynamic SGS models (Sarwar *et al.* 2017), prompting extensive research efforts to optimize its design and implementation (Germano 1986b, c, Najjar & Tafti 1996, Sagaut & Grohens 1999, Pruett & Adams 2000, Berland *et al.* 2011).

A prominent LES strategy that leverages this idea is the Approximate Deconvolution (AD) method (Stolz & Adams 1999), involving a two-stage process: the soft deconvolution, which deals with the SFS reconstruction (up to the Nyquist wavenumber) from the solution using an (approximate) inverse explicit filter, and the hard deconvolution—also referred to as (secondary)

regularization— which addresses SGS modeling. Several characteristics are known to be desirable for explicit filters used for AD-LES, including no alteration at low to moderate wave numbers, a large attenuation near the Nyquist, a sharp roll-off, a low commutation error, a small dispersion error, no amplification at any wavenumber (filter stability), no transfer function overshoot, the positivity, the smoothness, a strong anti-aliasing, etc. (Vasilyev *et al.* 1998, Sagaut 2006, Layton & Rebholz 2012, Najafiyazdi *et al.* 2023). Originally, Stolz and Adams (1999) used the implicit Padé filters for AD-LES. In a pioneering work, Vasilyev *et al.* (1998) introduced criteria in terms of filter moments to achieve a desired order of the commutation error. They proposed discrete filters in computational space, imposing these criteria along with additional constraints to adjust a prespecified filter width and ensure perfect attenuation at the Nyquist. Stolz et al. (2001) extended this method by defining discrete filters in physical space. They enforced a higher order of commutation error at the boundaries, compared to internal nodes. They also used an additional constraint to minimize the dispersion error (the imaginary part of the filter transfer function). They showed that defining filter moment criteria in physical space is superior to the original formulation in computational space (Vasilyev *et al.* 1998), since the former is computationally more efficient and conditions for a small alteration at low-wavenumbers are automatically met for the filters with an explicit rule. In addition, they designed new (implicit-form) Padé filters to meet all these criteria. All these filters were restricted to structured grids.

Early attempts to design explicit filters for unstructured grids date back to the work by Marsden *et al.* (2002) and Haselbacher and Vasilyev (2003). They developed 2[nd]-order filters for unstructured grids through a local neighbor nodes selection procedure with a 3-node triangular stencil in 2D or 4-node tetrahedral stencil in 3D. The first group (Marsden *et al.* 2002) employed linear combinations of polynomial interpolation functions, and the second group (Haselbacher & Vasilyev 2003) adopted the least-squares gradient reconstruction method. However, these filter's properties were sensitive to grid node distribution. In addition, constructing geometric simplices— especially in 3D or near boundaries—added the complexity and declined the stability (especially in the case of skewed or stretched elements) in these methods.

The differential filtering, originally proposed by Germano (1986a, c), is an alternative solution for unstructured grids. San *et al.* (2015) compared the performance of the Padé, box, and elliptic differential filters for AD-LES of Taylor-Green vortex and pointed out that hyper-differential filters perform better than the others. However, in their subsequent studies (San 2016, Maulik &

San 2018) on filter types, including Padé, hyper-differential, and discrete binomial, for decaying Burgers turbulence, they concluded that the Padé filter with $\alpha = 0.3$ outperforms the others. Chang et al. (2022) studied some invertible filter types, including Gaussian, differential, hyper-differential, Chebyshev, etc., for the approximate and direct deconvolution methods in LES of isotropic turbulence. They reported that differential filters need a larger filter-to-grid size ratio compared to the others to give rise to accurate reconstructions.

Differential filters do not possess some characteristics desired for AD-LES, e.g., they lose commutativity, do not have a large attenuation at the Nyquist, and are rather dissipative for low to moderate wave numbers. Recently, Najafi-Yazdi et al. (2015, 2023) tackled the latter two issues by designing an improved differential filter in the context of the finite element discretization with several specific element topologies and shape functions. The incorporation of these filters into Finite Volume Method (FVM) solvers, or their application with general grid topologies or discretization schemes, needs further investigation. In addition, these filters introduce challenges at non-periodic boundaries, such as walls, where the velocity gradient must be provided or numerically reconstructed (Bose 2012), which produces an additional uncertainty. Therefore, despite the wide usage of differential filters (Vreman 2004, Nouri et al. 2011, Aljure et al. 2014, López Castaño et al. 2019, Saeedipour et al. 2019, Schneiderbauer & Saeedipour 2019, Rauchenzauner 2022, Rauchenzauner & Schneiderbauer 2022), the lack of commutativity, a high computational cost, a poor parallelizability, and the introduction of errors due to uncertainties in the numerical treatment of the differential filter at wall boundaries are the general drawbacks of these filters, yet to be addressed.

Other important studies on the design of explicit filters include the work by Kim et al. (2021), who improved the parallel efficiency via a recursive filtering approach, replacing wide-stencil filters with repeated narrow-stencil ones. This reduced memory usage and inter-process communication but increased arithmetic operations by ~25% and remained limited to structured grids. The divergence-preserving property of filters was highlighted by Agdestein and Sanderse (2025). They proposed a filter on unstructured grids using face averaging, rather than volume averaging, along with grid cell agglomeration.

Based on our literature review, the design and implementation of efficient explicit filters specifically tailored for FVM-based solvers on general unstructured grids remain underexplored. Here, we aim to design a new explicit filter for AD-LES on unstructured grids (or non-uniform structured grids with high aspect ratios), partially addressing the deficiencies of existing filters. Most notably, while recursive filters have shown promise in the structured grid context (Kim *et al.* 2021), their adaptation to unstructured grids and their optimization for accuracy and stability remain unexplored. This study is conducted towards filling these gaps by proposing a novel recursive filter on unstructured grids leveraging a multi-objective optimization framework to fine-tune filter coefficients for desirable properties.

The remainder of this paper is structured as follows: Section 2 provides the mathematical modeling of AD-LES, benchmarks used for *a posteriori* tests, and numerical methods. Section 3 discusses explicit filters for AD-LES, outlining the filter design criteria, analyzing the limitations of existing filtering approaches, and introducing a novel recursive filter that addresses these challenges while meeting the desired properties. Section 4 features the results of the *a posteriori* analyses that rigorously evaluate the new filter performance, in comparison to the widely used filters for general unstructured grids. Finally, section 5 summarizes the key findings and overarching conclusions of this study.

## 2. AD-LES modelling

### 2.1. Mathematical modelling

The Navier-Stokes (NS) equations governing incompressible flows of a Newtonian fluid are expressed as:

$$\partial_t u_i + \partial_j(u_i u_j) + \partial_i p - \partial_j(2\nu S_{ij}) \equiv \partial_t u_i + \mathcal{NS}(u_i) = 0, \tag{1}$$

$$\partial_i u_i = 0, \tag{2}$$

where $\partial_t$ and $\partial_i$ denote the time ($\partial/\partial t$) and spatial ($\partial/\partial x_i$) derivatives, respectively. These equations include the velocity vector, $u_i$, the kinematic pressure, $p$, the kinematic viscosity, $\nu$, and the strain-rate tensor, $S_{ij}$, defined as:

$$S_{ij} = \frac{1}{2}(\partial_j u_i + \partial_i u_j). \tag{3}$$

The term "kinematic" refers to the division of the corresponding variable by the constant fluid density, $\rho$. For brevity, the term "kinematic" is omitted hereafter and assumed implicit for all pressures, viscosities, and stresses.

The filtered NS equations solved in an AD-LES approach can be written as:

$$\partial_t \bar{\tilde{u}}_i + \partial_j(\bar{\tilde{u}}_i \bar{\tilde{u}}_j) = -\partial_i \bar{\tilde{p}} + \partial_j(2\nu \bar{\tilde{S}}_{ij}) - \partial_j T_{ij}, \qquad (4)$$

$$\partial_i \bar{\tilde{u}}_i = 0, \qquad (5)$$

where $\widetilde{(.)}$ denotes the grid filter, $\overline{(.)}$ or $G*(.)$ the explicit filter where $G$ represents the explicit filter kernel, and $T_{ij}$ is the Sub-Filter-Scale (SFS) stress tensor defined by:

$$T_{ij} \equiv \overline{\widetilde{u_i u_j}} - \bar{\tilde{u}}_i \bar{\tilde{u}}_j = b_{ij} + a_{ij}. \qquad (6)$$

To solve Eqs. (4) and (5) for $\bar{\tilde{u}}_i$ and $\bar{\tilde{p}}$, a closure model is required for the unclosed SFS stress, $T_{ij}$, (or $\partial_j T_{ij}$). This quantity is decomposed into two parts, the deconvolved SFS stress, $b_{ij}$, and the modeled SFS stress, $a_{ij}$. A widely-used decomposition of $T_{ij}$ can be written as (Carati *et al.* 2001):

$$b_{ij} = \overline{\tilde{u}_i \tilde{u}_j} - \bar{\tilde{u}}_i \bar{\tilde{u}}_j, \qquad (7)$$

$$a_{ij} = \overline{\widetilde{u_i u_j}} - \overline{\tilde{u}_i \tilde{u}_j} = \bar{\tau}_{ij}, \qquad (8)$$

where $\tau_{ij}$ is the SGS stress tensor defined by:

$$\tau_{ij} = \widetilde{u_i u_j} - \tilde{u}_i \tilde{u}_j. \qquad (9)$$

Here, the deconvolved SFS stress is reconstructed by the Scale-Similarity ADM (SSADM) soft deconvolution (Stolz & Adams 1999):

$$b_{ij}^M = \overline{u_i^\star u_j^\star} - \overline{u_i^\star}\,\overline{u_j^\star}, \qquad (10)$$

where the superscript $M$ denotes the modeled counterpart of a quantity, and $u_i^\star$ is the (deconvolved) approximation to $\tilde{u}_i$. We use the Van Cittert iterative algorithm (Van Cittert 1931) with five iterations for the approximate deconvolution operation. The modeled SFS stress, $a_{ij}$, encapsulates the SGS effect and must be modeled ($a_{ij}^M$) using an SGS closure or (secondary) regularization. For this purpose, the mixed Alternative Linear Dynamic Model with Equilibrium assumption (ALDME) model proposed by Amani *et al.* (2024) is employed, where

$$a_{ij}^{M,r} = \bar{\tau}_{ij}^{M,r} = G * \tau_{ij}^{M,r}, \qquad (11)$$

$$-\tau_{ij}^{M,r} = 2(C_D \widetilde{\Delta}^2 S^\star) S_{ij}^\star = 2\nu_r S_{ij}^\star, \qquad (12)$$

where superscript $r$ refers to the deviatoric part of a tensor, $S^\star = (2 S_{ij}^\star S_{ij}^\star)^{1/2}$, $\widetilde{\Delta}$ the grid filter width, $\nu_r$ the SGS eddy-viscosity, and $C_D$ the dynamic model coefficient computed by:

$$C_D = \frac{L_{ij}^r M_{ji}}{M_{mn} M_{nm}}, \qquad (13)$$

$$L_{ij} = \widetilde{u_i^\star u_j^\star} - \widetilde{u_i^\star}\,\widetilde{u_j^\star}, \qquad (14)$$

$$M_{ij} = -8\widetilde{\Delta}^2 \widetilde{S^\star S_{ij}^\star}, \tag{15}$$

and $\widetilde{(.)}$ is a test box filter of width $\breve{\Delta}= 2\tilde{\Delta}$.

The flow statistics are computed using the ensemble (Reynolds) averaging. Key quantities include the mean velocity $\langle u_i \rangle$ and the Reynolds stress, $\langle u_i' u_j' \rangle$, where the fluctuating velocity is $u_i' = u_i - \langle u_i \rangle$ and $\langle \ \rangle$ denotes the ensemble averaging operator. In the context of AD-LES, these quantities are approximated under the assumption $\langle q \rangle \approx \langle \bar{\tilde{q}} \rangle$ (Pope 2000), as follows:

$$\langle u_i \rangle \approx \langle \bar{\tilde{u}}_i \rangle; \ \langle S_{ij} \rangle \approx \langle \bar{\tilde{S}}_{ij} \rangle, \tag{16}$$

$$\langle u_i' u_j' \rangle \approx \langle (\bar{\tilde{u}}_i - \langle \bar{\tilde{u}}_i \rangle)(\bar{\tilde{u}}_j - \langle \bar{\tilde{u}}_j \rangle) \rangle + \langle b_{ij} \rangle + \langle a_{ij}^r \rangle + \frac{2}{3} \langle \overline{k_{SGS}} \rangle \delta_{ij}, \tag{17}$$

where $k_{SGS}$ represents the SGS Turbulent Kinetic Energy (TKE). For $k_{SGS}$, the following estimation can be employed (Sullivan et al. 2003):

$$k_{SGS} \approx \left(\frac{C_D}{C_e}\right)^{\frac{2}{3}} \tilde{\Delta}^2 S^{\star 2}, \tag{18}$$

where $C_e = 1$ represents the model constant.

## 2.2. The a posteriori benchmarks

### 2.2.1. Turbulent channel flow

The fully-developed turbulent flow of a Newtonian fluid between two parallel infinite plates, driven by a constant mean pressure gradient—commonly termed "channel flow"—is a benchmark problem frequently employed for both *a posteriori* and *a priori* analyses. The flow dynamics are characterized by a single dimensionless parameter, the friction Reynolds number, defined as $Re_\tau = u_\tau H/\nu$, where $H$ is the half-channel height, $u_\tau = \tau_w^{1/2}$ the friction velocity, and $\tau_w$ the wall shear stress. In this study, the Direct Numerical Simulation (DNS) dataset by Moser *et al.* (1999) is used as the reference solution. The computational domain, illustrated in figure 1a, is configured with dimensions of $(L \times 2H \times W) = (2\pi \times 2 \times \pi)$. The results are presented as a function of the normalized wall distance, $y^+ = y u_\tau/\nu$, where $y$ represents the distance from the wall.

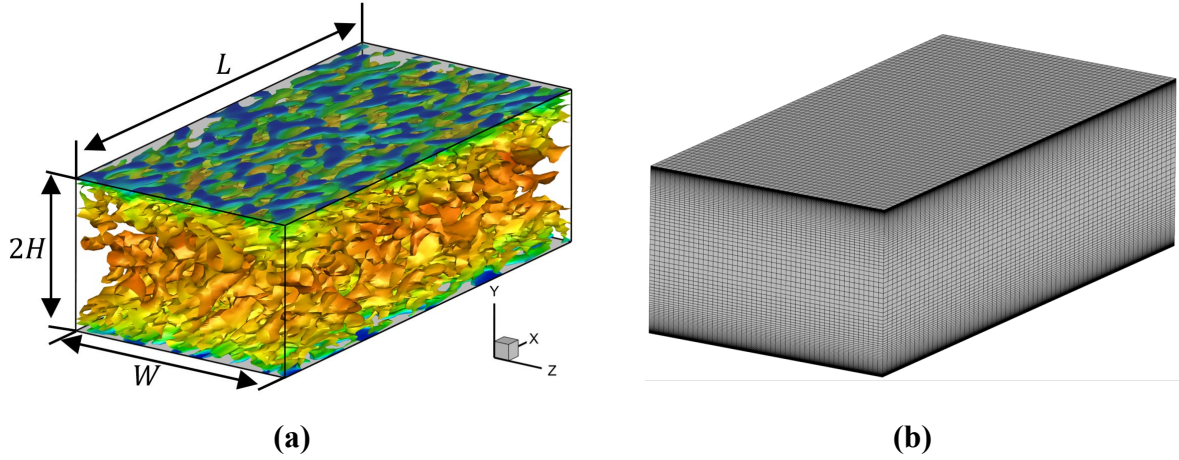

Figure 1: a) A schematic geometry of the channel flow and Q-criterion iso-surfaces predicted by the present AD-LES, and b) a computational grid used for simulations. ($Re_\tau = 395$).

### 2.2.2. Taylor-Green vortex

The 3D Taylor-Green Vortex (TGV) is a canonical benchmark, extensively used to investigate vortex dynamics. This problem involves the decay of a 3D field of vortices within a cubic domain, subject to periodic boundary conditions on all faces. The computational domain, illustrated in figure 2, is a cube with a side length of $2\pi L$, where $L$ serves as the reference length. The initial flow field at $t = 0$ comprises a distribution of distinct large-scale vortical structures defined by the following velocity field (DeBonis 2013):

$$u = V_0 \sin\left(\frac{x}{L}\right) \cos\left(\frac{y}{L}\right) \cos\left(\frac{z}{L}\right),$$
$$v = -V_0 \cos\left(\frac{x}{L}\right) \sin\left(\frac{y}{L}\right) \cos\left(\frac{z}{L}\right), \qquad (19)$$
$$w = 0,$$

where $V_0$ denotes the initial velocity amplitude. The corresponding initial pressure field is analytically obtained to satisfy the Poisson's equation for incompressible flows as:

$$p = p_0 + \frac{\rho_0 V_0^2}{16}\left[\cos\left(\frac{2x}{L}\right) + \cos\left(\frac{2y}{L}\right)\right]\left[\cos\left(\frac{2z}{L}\right) + 2\right] \qquad (20)$$

where $p_0$ and $\rho_0$ are the reference pressure and density, respectively. The flow dynamics is characterized by the Reynolds number, $Re = \rho_0 V_0 L/\mu$. The reference values are chosen according to (DeBonis 2013) as: $L = 0.001524\ m$, $V_0 = 34.396\ m/s$, $p_0 = 0$ Pa, $\rho_0 = 1.0\ kgm^{-3}$, and $\mu = 3.2762 \times 10^{-5}\ kgm^{-1}s^{-1}$, leading to $Re = 1600$. A reference DNS solution using a high-order (4th-order in space and 3rd-order in time) finite-difference discretization on uniform structured grids of 512 × 512 × 512 cubic cells is available (DeBonis 2013).

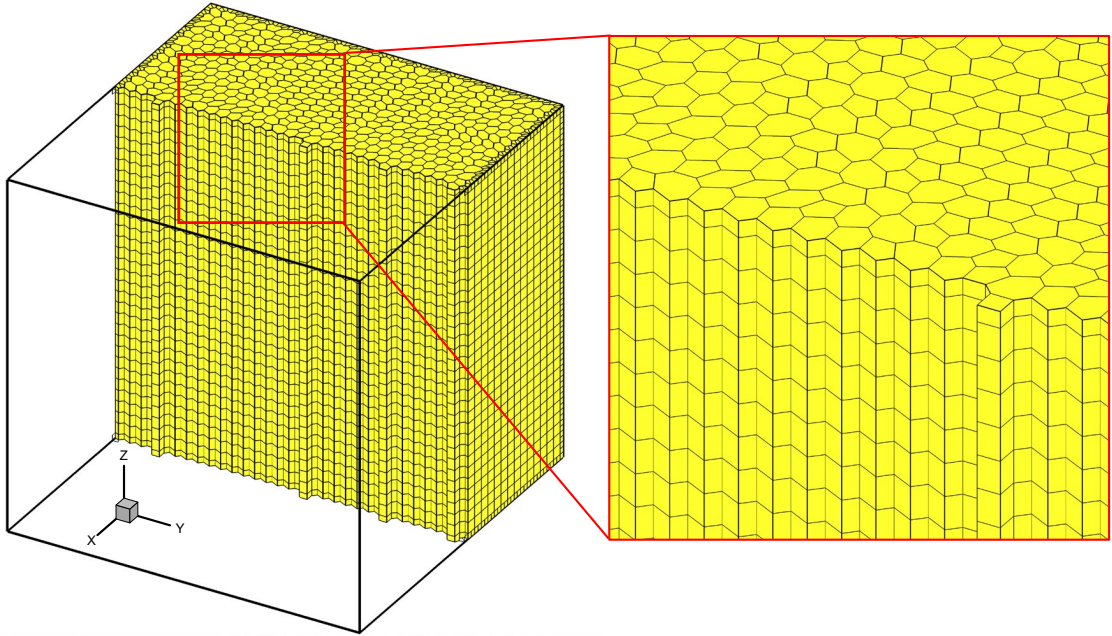

Figure 2: The computational domain and unstructured grid composed of prism cells, used for the current LES of TGV benchmark.

*2.3. Numerical methods*

In this study, the cell-centered collocated FVM implemented for general unstructured grids in the OpenFOAM ESI CFD package (www.openfoam.com), version 1912, was enhanced to incorporate ADM-type closures and the new filtering operation. For the present LES of channel flow, the pressure-velocity coupling is resolved using the Pressure Implicit with Splitting of Operators (PISO) algorithm (Issa 1986) employing two pressure-correction loops. Temporal discretization is achieved using the implicit second-order backward scheme (Greenshields 2015, Moukalled et al. 2016). Gradients are computed via the Green-Gauss cell-based method, utilizing the "Gauss linear" scheme (Greenshields 2015). The advection term in the momentum equation is discretized using the second-order pure-centered "Gauss linear" scheme, while variable interpolation to cell faces is performed using the "linear" scheme. For the pressure equation, the resulting system of linear algebraic equations is solved using the Preconditioned Conjugate Gradient (PCG) method, with the Diagonal Incomplete-Cholesky (DIC) preconditioner combined with Gauss-Seidel smoothing (Saad 2003). The momentum equation is solved using the Stabilized Preconditioned Bi-Conjugate Gradient (PBiCGStab) solver, preconditioned with the Diagonal Incomplete-LU (DILU) method (Van der Vorst 1992, Barrett et al. 1994, Saad 2003). A normalized residual tolerance of $10^{-6}$ is set as the convergence criterion for all variables at each time step. To ensure numerical stability and accuracy, the time step is dynamically adjusted to maintain a maximum Courant-Friedrichs-Lewy (CFL) number of 0.5.

The computational domain is depicted in figure 1a. Periodic boundary conditions are applied to the streamwise and spanwise boundaries for all flow variables, with the mean pressure gradient enforced through an additional source term in the momentum equation. At the lower and upper walls, the no-slip condition is imposed on the velocity field, while a zero-gradient condition is applied to the pressure. The grid is uniformly distributed in the streamwise and spanwise directions, with wall-normal refinement achieved using a hyperbolic tangent function to cluster points near the walls. Details of the LES grid, along with the reference DNS grid (Moser et al. 1999), are provided in table 1. As can be seen in this table, the grid has a high aspect ratio which is typical of boundary layer grids used for LES and DNS. The parameters $(\Delta y^+)_w$ and $(\Delta y^+)_c$ in table 1 refer to the grid cell size (in the y direction) on the wall and at the channel center plane, respectively. The computational grid topology is illustrated in figure 1b.

Table 1: A comparison of the present LES and reference DNS (Moser et al. 1999) grids.

| Solution | $\Delta x^+$ | $[(\Delta y^+)_w, (\Delta y^+)_c]$ | $\Delta z^+$ | $\widetilde{\Delta}/h_{DNS}$ | $N_x \times N_y \times N_z$ |
|---|---|---|---|---|---|
| **DNS ($Re_\tau = 395$)** | 10 | [0.03, 6.5] | 6.5 | 1 | $256 \times 193 \times 192 = 9,486,336$ |
| **LES ($Re_\tau = 395$)** | 38.78 | [0.82, 25] | 25.85 | 3.90 | $64 \times 110 \times 48 = 337,920$ |
| **DNS ($Re_\tau = 590$)** | 9.7 | [0.044, 7.2] | 4.8 | 1 | $384 \times 257 \times 384 = 37,896,192$ |
| **LES ($Re_\tau = 590$)** | 40 | [0.8, 35.4] | 25 | 4.72 | $75 \times 132 \times 94 = 930,600$ |

To accelerate the simulations, the Wall-Adapting Local Eddy-viscosity (WALE) model (Nicoud & Ducros 1999) is utilized for generating an initial solution for the AD-LES, due to its robustness and simplicity. The velocity field initialization procedure and the solution steps to achieve a WALE model solution have been detailed elsewhere (Amani et al. 2023, Taghvaei & Amani 2023). Starting from the WALE solution, the simulations continue for 30 $FTT$, where $FTT$ is the Flow-Through-Time ($FTT = L/U_b$) and $U_b$ is the channel bulk velocity, marking the attainment of a statistically stationary condition. Subsequently, the simulations continue to collect the statistics by the ensemble averaging, ⟨ ⟩, taken here as a spatial averaging in the two homogeneous directions ($x$ and $z$) and a time-averaging for 70 $FTT$. Concerning the validation of the unstructured-grid finite volume solver, i.e., OpenFOAM, the present solver has been extensively validated for LES of turbulent flows in references (Tofighian *et al.* 2019, Amani *et al.* 2023, Taghvaei & Amani 2023).

Diverging solutions were observed for conventional pure Dynamic Eddy-Viscosity (DEV) and mixed AD-DEV models, see also (Amani *et al.* 2024). As a potential solution, the clipping method based on the realizability conditions proposed by Mokhtarpoor and Heinz (2017) was implemented, but it was not sufficient to resolve the instability issue for many cases. Subsequently, the Positive Total Viscosity (PTV) approach was employed as $\nu_r + \nu \geq 0$ or $\nu_r/\nu \geq -1$, which successfully addressed the stability concerns.

For the TGV benchmark, the numerical setup is modified to address the specific stability requirements of the transitional vortex breakdown. The pressure-velocity coupling within the PISO algorithm is augmented to include two non-orthogonal corrector steps to enhance accuracy on unstructured computational grids. Regarding spatial discretization, the gradient of the velocity field is computed using a cell-limited Gauss linear scheme (Guo & Wang 2025) with a limiting coefficient of 1.0 to mitigate numerical oscillations. The advection term in the momentum equation employs the Linear-Upwind Stabilized Transport (LUST) scheme (Cao & Tamura 2016)—a fixed blend of linear and linear-upwind schemes—replacing the pure-centered approach used for the channel flow benchmark to optimally balance dissipation and dispersion errors. Furthermore, to capture the rapid temporal evolution of the small-scale structures, the time step is dynamically restricted to maintain a maximum CFL number of 0.1. The ensemble averaging, $\langle \ \rangle$, for the TGV problem is taken as the volume-averaging within the whole computational domain (DeBonis 2013).

## 3. Explicit filters for AD-LES

### *3.1. The filter design criteria*

A 3D discrete explicit filter can be expressed as a weighted sum of values at cell centers, as follows (Sagaut & Grohens 1999):

$$(G * \phi)_i = \bar{\phi}_i = \sum_{j=1}^{N_c} a_{i,j} \phi_j, \tag{21}$$

where $\bar{\phi}_i$ represents the filtered variable at cell $i$, the summation is over all grid cells, and $a_{i,j}$ denotes the weighting (or filter) coefficient associated with cells $i$ and $j$. The discrete filter's transfer function in wavenumber space is defined by (Sagaut & Grohens 1999):

$$\hat{G}(\boldsymbol{k}) = \sum_{j=1}^{N_c} a_{i,j} e^{-i\boldsymbol{k}\cdot(\boldsymbol{x}_i-\boldsymbol{x}_j)}, \tag{22}$$

where $\boldsymbol{x}_i$ is the position vector (of the center) of cell $i$ and $\boldsymbol{k}$ is the wavenumber vector. The bold symbols indicate vector quantities, $\widehat{(.)}$ the complex numbers, and $i = \sqrt{-1}$. The dimensionless wavenumber can be defined by $\boldsymbol{\omega} = (k_x h_x, k_y h_y, k_z h_z)$ (Sengupta & Bhumkar 2010), where $h_i$ is the characteristic grid spacing in the $i^{\text{th}}$-direction, thus:

$$\boldsymbol{k} = \left(\frac{\omega_x}{h_x}, \frac{\omega_y}{h_y}, \frac{\omega_z}{h_z}\right). \tag{23}$$

For AD filters, several desirable properties have been extensively discussed in the literature (Vasilyev *et al.* 1998, Sagaut 2006, Layton & Rebholz 2012, Najafiyazdi *et al.* 2023) and are outlined as follows:

1. Normalization: The filter should not alter a uniform field,

$$\left|\hat{G}(\boldsymbol{\omega}=0)\right| = 1 \Rightarrow \sum_{j=1}^{N_c} a_{i,j} = 1. \tag{24}$$

2. High-wavenumber attenuation: Vanishing the filter transfer function at the Nyquist wavenumber ($|\boldsymbol{\omega}| = \pi$),

$$\left|\hat{G}(|\boldsymbol{\omega}| = \pi)\right| = 0. \tag{25}$$

3. Preadjusted filter width: The filter transfer function must attain a value of 0.5 at the specified cutoff frequency $|\boldsymbol{\omega}| = \omega_c$,

$$\left|\hat{G}(|\boldsymbol{\omega}| = \omega_c)\right| = 0.5. \tag{26}$$

4. Commutativity: For non-uniform computational grids, it is essential to preserve the commutativity property, the communication of the filter and difference operators, to a reasonable degree. To achieve $O(\Delta^K)$ accuracy, the filter's moments (up to order $K - 1$) in the physical space must be zero. For a second-order accuracy ($K = 2$) in FVM, the (dimensionless) first moments must vanish;

$$\begin{aligned} M_x^1 &= \frac{1}{h_x}\sum_{j=1}^{N_c} a_{i,j}(x_i - x_j) = 0, \\ M_y^1 &= \frac{1}{h_y}\sum_{j=1}^{N_c} a_{i,j}(y_i - y_j) = 0, \\ M_z^1 &= \frac{1}{h_z}\sum_{j=1}^{N_c} a_{i,j}(z_i - z_j) = 0. \end{aligned} \tag{27}$$

5. A small dispersion error: For this goal, it is necessary to minimize the imaginary part of the transfer function,

$$Im\{\hat{G}(\boldsymbol{\omega})\} \to 0; \forall \boldsymbol{\omega}. \tag{28}$$

6. Positivity: The real part of the transfer function must be strictly positive to ensure physical consistency and numerical stability,

$$Re\{\hat{G}(\boldsymbol{\omega})\} > 0; \forall \boldsymbol{\omega}. \tag{29}$$

7. No amplification (filter stability): The transfer function magnitude must be less than or equal unity to avoid artificial energy addition and buildup by filtering,

$$|\hat{G}(\boldsymbol{\omega})| \leq 1; \forall \boldsymbol{\omega}. \tag{30}$$

### 3.2. The necessity for a new FVM filter

A class of filters in the context of the FVM for unstructured grids includes differential filters. The application of differential filters in LES originated with the seminal work by Germano (Germano 1986b, c) who introduced a linear elliptic differential filter. Building on this foundation, different implicit and explicit forms of differential filters have been proposed, the former demands a high computational cost (Bose 2012). Here, a widely-used explicit differential filter variant, known as Laplace filter, also denoted here as "laplaceFilter", is considered (Weller *et al.* 1998):

$$\bar{\phi} = \phi + \boldsymbol{\nabla} \cdot (\alpha \boldsymbol{\nabla} \phi), \tag{31}$$

where $\alpha$ is an adjustable coefficient that determines the filter width. An FVM discretized form of the Laplace filter is expressed as:

$$\bar{\phi}_i = \phi_i + \frac{1}{V_i} \sum_{f=1}^{N_{f,i}} [(\alpha)^f (\boldsymbol{\nabla}\phi)^f \cdot \boldsymbol{S}_f], \tag{32}$$

where the sum is over all faces ($f$) of cell $i$, $V$ is the cell volume, $\alpha = V^{\frac{2}{3}}/c$ is a field assigned to cell centers, $(.)^f$ is the interpolation of a quantity at face $f$, $\boldsymbol{S}_f$ is the area vector of face $f$, and $c$ is a user-defined constant determining the filter width. A value of $c = 24$ corresponds to a grid filter ($\Delta = h$), while a value of $c = 6$ approximates a filter width of $\Delta = 2h$.

For wall cell faces, the determination of the velocity field gradient—which is required in Eq. (32)—is not straightforward. The velocity boundary condition at walls is a zero-valued Dirichlet condition. Consequently, the gradient cannot be directly specified and should be either extrapolated or estimated through *ad hoc* relations. This introduces inherent uncertainties, which can compromise the accuracy and robustness of simulations using the Laplace Filter in the vicinity of solid boundaries. This issue becomes particularly pronounced in AD-LES with algorithms like the Van Cittert method, where the filter is applied iteratively.

More importantly, our channel-flow *a posteriori* analyses (in section 4) demonstrate that the Laplace filter exhibits strong instabilities on non-uniform grids, resulting in divergence of the solution. To explain the reason for this issue, the 3D instantaneous axial velocity field of a channel flow at $Re_\tau = 395$, obtained from an LES solution on a typical channel-flow non-uniform Cartesian grid (with a maximum aspect ratio of 50 and wall-normal growth rate of 1.1), is considered here as the input ($u$) to the 3D explicit filtering operation. Figure 3a presents the velocity profile at a channel cross-section along with the corresponding filtered ($\bar{u}$) and deconvoluted ($u^\star$) fields profiles, using a Laplace filter of width $\varDelta = 2h$ and the Van-Cittert deconvolution algorithm. The deviation $(u^\star - u)/u$ is plotted in figure 3b for various Van-Cittert iterations ($N$). The filter is clearly unstable near the wall, as evidenced by the pronounced oscillations in both the filtered and deconvoluted profiles (figure 3a). As shown in figure 3b, increasing the number of Van-Cittert iterations intensifies these oscillations and propagates them further towards the channel center. Note that a zero-valued velocity gradient at the walls, which is commonly used for this filter, was adopted for the results in figure 3, however, an extrapolated velocity gradient at the boundaries was also tested, which resulted in the same issue. To further support that these instabilities are mainly originated from the grid configuration rather than the uncertainty of the Laplace filter at the boundaries, the same velocity profile ($u$) is mapped onto a uniform Cartesian grid with a unity aspect ratio, and the computed filtered and deconvoluted profiles on this grid are shown in figure 3c. The results demonstrate a stable filter performance, with increasing iterations reducing the deviation between $u^\star$ and $u$ (figure 3d). This critical issue of the Laplace filter under non-uniform grid conditions will be further analyzed based on the properties of filters in the current section.

Another limitation of the standard Laplace filter formulation lies in the definition of the diffusion coefficient $\alpha$. As shown in Eq. (32), while $\alpha$ changes locally with the grid size, it is treated isotropically as a scalar. This isotropic definition implies that the filter applies the same diffusion coefficient across all faces, regardless of the cell aspect ratio. On highly anisotropic grids, e.g., boundary layers, this inability to adapt to the grid anisotropy results in excessive filtering or stability issues in the wall-normal direction.

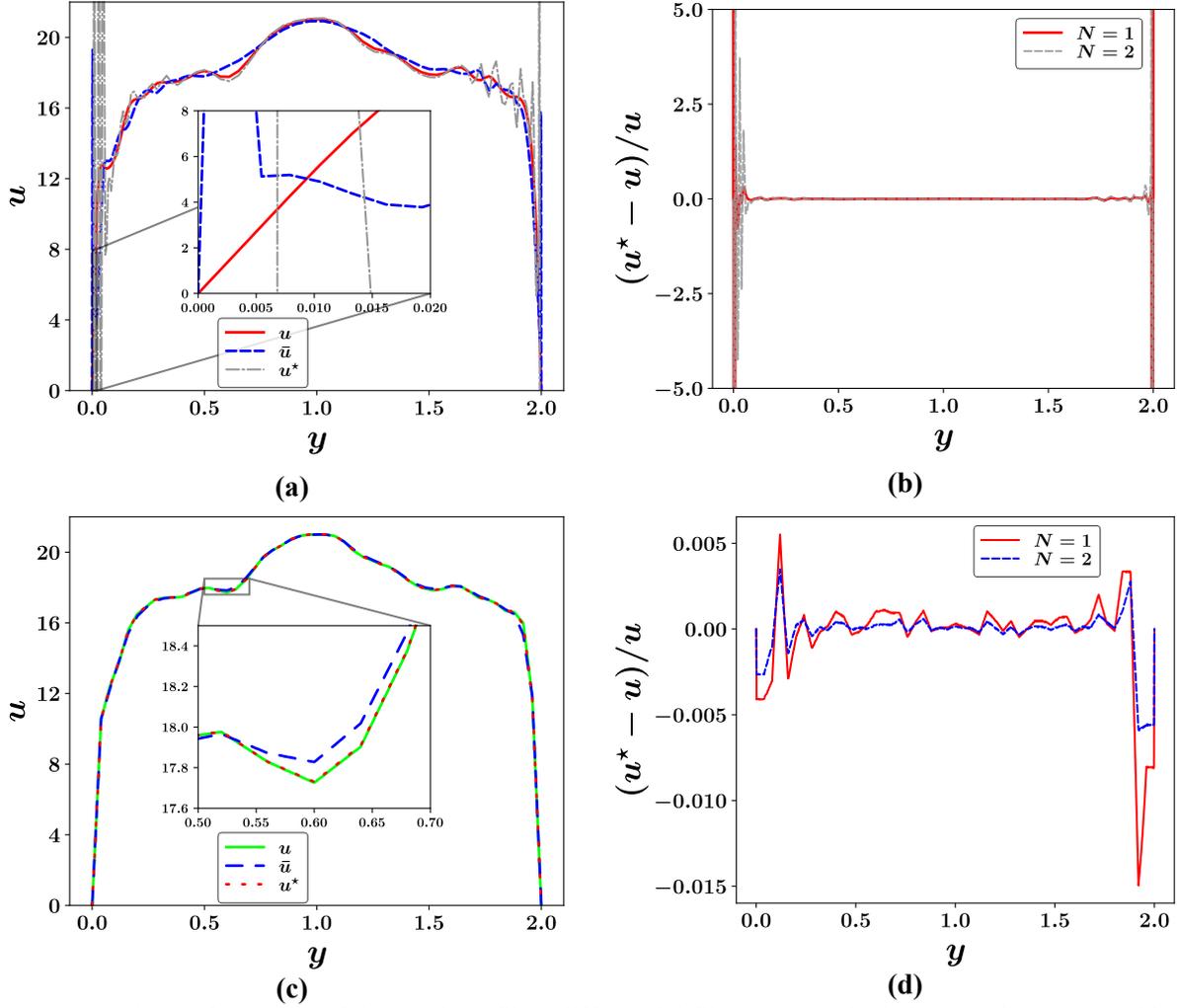

**Figure 3:** The performance of the Laplace filter differential filter on a typical channel-flow non-uniform grid (a, b) and on a uniform grid (c, d): The original ($u$), filtered ($\bar{u}$), and deconvoluted ($u^\star$) axial velocity profiles at a channel cross-section for $N = 2$ (a, c), and the normalized deviation of the deconvoluted velocity from the original velocity for different $N$s (b, d).

Another discrete filtering approach applicable to general unstructured grids is the use of face-averaging methods. A key advantage of this technique is that it does not require the additional treatment of the velocity gradient at the boundaries. Furthermore, it is computationally efficient, being approximately five times faster than the Laplace filter when applied to the velocity field in a channel flow (based on our computational cost analysis). Among face-averaging methods, the area-weighted face average, known as "simpleFilter", is the most widely-used, which is defined by:

$$\bar{\phi}_i = \frac{\sum_{f=1}^{N_{f,i}}[(\phi)^f |\boldsymbol{S}_f|]}{\sum_{f=1}^{N_{f,i}}|\boldsymbol{S}_f|} \equiv F_S(\phi_i). \tag{33}$$

To analyze the properties of this filter for a canonical test case, a 3D uniform Cartesian grid with a geometric aspect ratio of $R$, where $h_x = h_z = Rh_y$, is considered. It can be shown that the incorporation of the Laplace and simple filters, i.e., Eq. (32) or (33), on this grid (and the use of the standard central $2^{nd}$-order discretization of $(\nabla \phi)^f$ for the former filter) leads to the expressions in the form of Eq. (21) for these filters with the coefficients, $a_{i,j}$, given in table 2. For the sake of comparison, the coefficients for a box filter of width 2 are also included. Additionally, the coefficients for the second-order Padé filter (Lele 1992) at a dimensionless cut-off wavenumber of $\pi/2$ are reported; the derivation is provided in Appendix A. A detailed comparison of the filter coefficients reveals several key points:

1. Aspect ratio dependence: The coefficients of both the Laplace and simple filters exhibit explicit dependence on the grid aspect ratio, i.e., parameter $R$. In contrast, the coefficients of the box and Padé filters are independent of the aspect ratio.

2. Directional variation: For the Laplace and simple filters, the coefficients in the y-direction are significantly different from those in the other directions—specifically, by a factor of $R^2$ in the case of the Laplace filter and $R$ in the case of the simple filter. This indicates that the filtering operation is highly anisotropic and predominantly applied in the y-direction. Conversely, the box and Padé filters maintain identical coefficients across all spatial directions.

3. Filter width control: The effective width of the Laplace and simple filters varies with the grid aspect ratio.

**Table 2:** Filter coefficients on a 3D uniform Cartesian grid with an aspect ratio of $R$ ($h_x = h_z = Rh_y$) for different filter types.

| Filter coefficients | Box filter | 2nd-order Padé filter | Laplace filter | simple filter | $F_M$ Filter |
|---|---|---|---|---|---|
| $a_{i,j,k}$ | $\frac{1}{8}$ | $\frac{1}{8}$ | $1 - \frac{R^{\frac{4}{3}}}{c}\left(\frac{4}{R^2} + 2\right)$ | $\frac{1}{2}$ | 0.5 |
| $a_{i\pm1,j,k}$ | $\frac{1}{16}$ | $\frac{1}{16}$ | $\frac{1}{cR^{\frac{2}{3}}}$ | $\frac{1}{4(2+R)}$ | $\frac{1}{12}$ |
| $a_{i,j\pm1,k}$ | $\frac{1}{16}$ | $\frac{1}{16}$ | $\frac{R^{\frac{4}{3}}}{c}$ | $\frac{R}{4(2+R)}$ | $\frac{1}{12}$ |
| $a_{i,j,k\pm1}$ | $\frac{1}{16}$ | $\frac{1}{16}$ | $\frac{1}{cR^{\frac{2}{3}}}$ | $\frac{1}{4(2+R)}$ | $\frac{1}{12}$ |
| $a_{i\pm1,j\pm1,k}$ | $\frac{1}{32}$ | $\frac{1}{32}$ | 0 | 0 | 0 |
| $a_{i\pm1,j,k\pm1}$ | $\frac{1}{32}$ | $\frac{1}{32}$ | 0 | 0 | 0 |
| $a_{i,j\pm1,k\pm1}$ | $\frac{1}{32}$ | $\frac{1}{32}$ | 0 | 0 | 0 |
| $a_{i\pm1,j\pm1,k\pm1}$ | $\frac{1}{64}$ | $\frac{1}{64}$ | 0 | 0 | 0 |

To provide a deeper insight into the characteristics of the Laplace and simple filters, we plot their transfer functions for four distinct computational grid configurations. Our analysis begins with uniform Cartesian grids. The transfer functions for the Laplace filter ($c = 6$) and simple filter under this condition can be computed by Eq. (22) using the coefficients, $a_{i,j}$, given in table 2. The magnitude, real part, and imaginary part of their transfer functions for a grid with an aspect ratio of unity are presented in figure 4a and b, respectively. Notably, neither filter exhibits a high attenuation at large wavenumbers. To examine the impact of grid anisotropy, figure 4c and d display the transfer functions for a uniform grid with an aspect ratio of $R = 50$. The simple filter shows a sufficient high-wavenumber attenuation only in the wall-normal ($y$) direction, but this comes at the expense of a significant reduction in attenuation in the other two directions. In the case of the Laplace filter (figure 4c), the presence of large negative values in the real part of its transfer function in the y-direction on the grid with a high aspect ratio indicates a violation of the positivity property (Eq. (29)). Likewise, the transfer function magnitude significantly exceeds unity, demonstrating the absence of the filter stability property (Eq. (30)). This well explains the instability of the filtering operation on the highly-anisotropic channel-flow grid seen in figure 3 and the divergence of the solution using this filter in our *a posteriori* tests (section 4). By the use of the filter stability criterion, it can be demonstrated that, for a uniform grid with an aspect ratio of $R$, the Laplace filter remains stable only when $R \leq (c/2)^{3/4}$ (the derivation is provided in Appendix B). For instance, with $c = 6$ ($\Delta = 2h$), the maximum permissible aspect ratio is 2.2795.

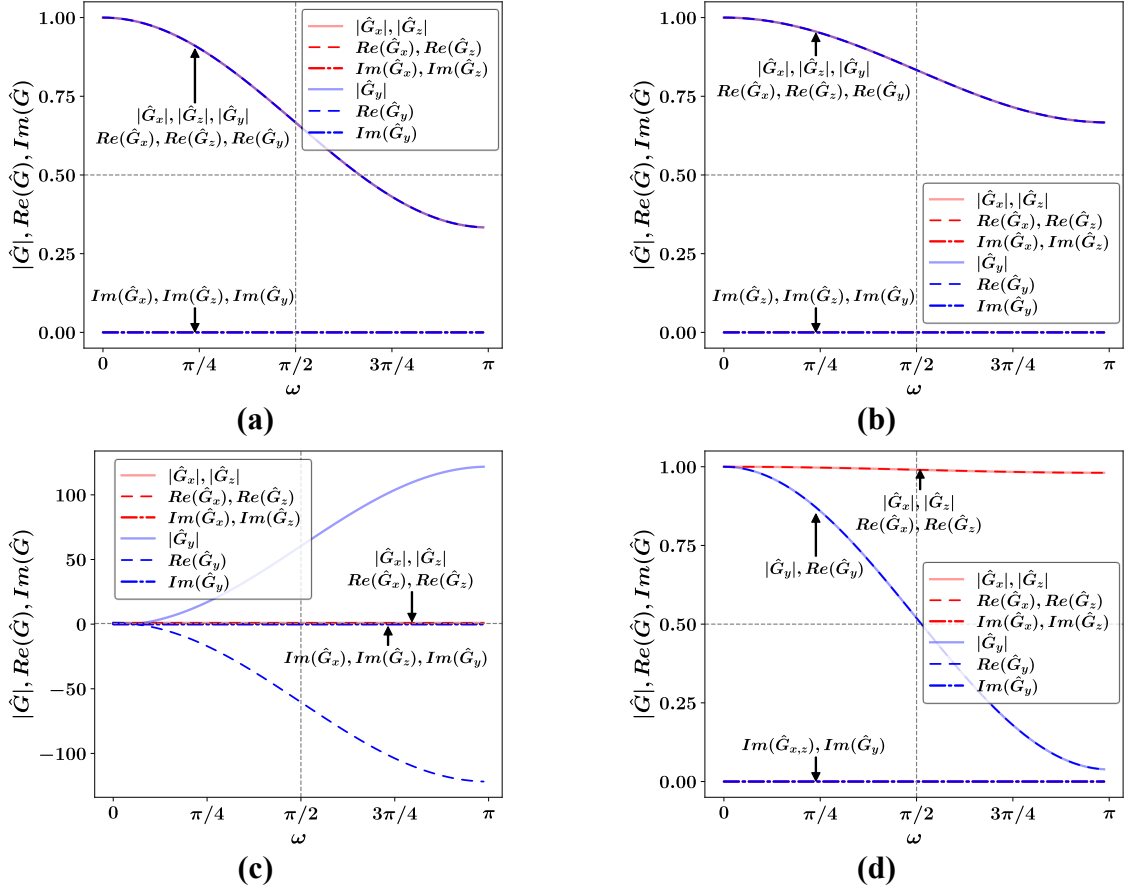

**Figure 4:** The effect of the grid aspect ratio on the magnitude, real part, and imaginary part of the filter transfer functions on uniform Cartesian grids: a) The Laplace filter $R = 1$, b) simple filter $R = 1$, c) Laplace filter $R = 50$, and d) simple filter $R = 50$. Another representation of the transfer functions in the range $R \in (1,50)$ is provided in figure 1 of Supplemental Material S1.

In the next step, we proceed to investigate the properties of the Laplace and simple filters under non-uniform Cartesian and unstructured grid conditions. Since an analytical determination of filter coefficients for these scenarios presents a high complexity, we overcome this challenge using MATLAB's (www.mathworks.com) Symbolic Math Toolbox. The code to compute the filter coefficients on these grids along with a description of its algorithm is given in Appendix C.

Using the computed filter coefficients on a non-uniform Cartesian grid with a growth rate of 1.1 in the y-direction and a unity aspect ratio, the transfer function (Eq. (22)) of Laplace and simple filters is illustrated in figure 5a and b, respectively. It can be observed that the real part and the magnitude of the transfer functions have not changed significantly compared to figure 4a and b. However, the imaginary part of the transfer function for both filters in the y-direction has become non-zero; for the Laplace filter, the maximum magnitude of the imaginary part is 0.05, and for the simple filter, it is 0.025. Therefore, the grid expansion produces a level of dispersion error (see Eq. (28)). The mutual effect of the growth rate and aspect ratio is examined in figure 5c and d, which respectively display the transfer function of the Laplace and simple filters for a non-uniform grid with a growth rate of 1.1 and an aspect ratio of 50 (a typical channel-flow grid). Again, the real part and the magnitude of the transfer functions have not changed significantly compared to the uniform grid in figure 4c and d, however, the imaginary part has become non-zero. With increasing the aspect ratio from 1 to 50, it is observed that the maximum magnitude of the imaginary part of the transfer function in the y-direction increases for both the Laplace and simple filters, reaching values of 9 and 0.08, respectively. This manifests that the mutual presence of the aspect ratio intensifies the impact of the growth rate. The first moments of the filters are also reported in table 3. Note that the simple filter is not commutative in the direction of the grid expansion ($M_y^1 \neq 0$) and its deviation from the commutativity property, Eq. (27), grows with the increase in the aspect ratio ($R$).

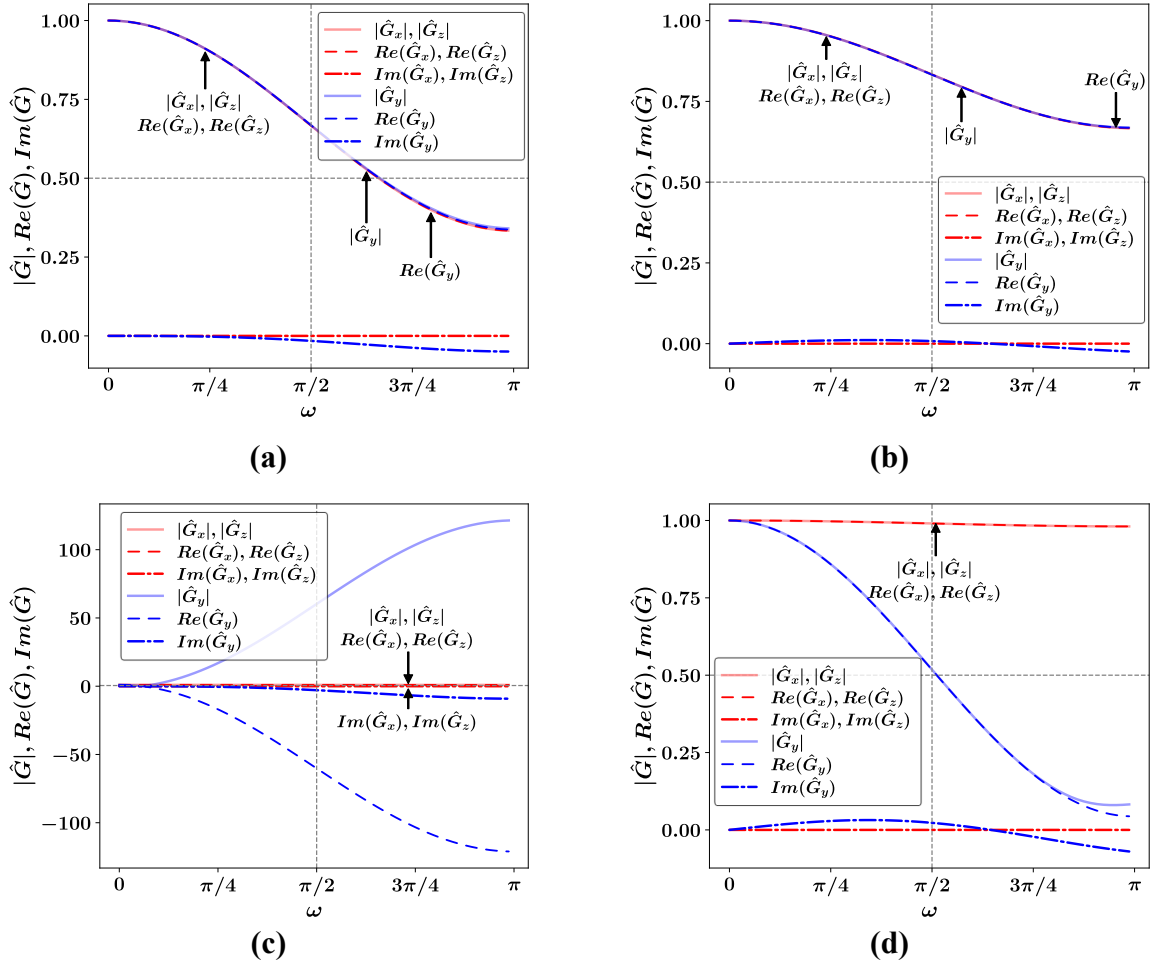

Figure 5: The effects of the grid growth rate ($GR$) and aspect ratio ($R$) on the magnitude, real part, and imaginary part of the filter transfer functions on non-uniform Cartesian grids: a) The Laplace filter ($GR = 1.1$ and $R = 1$), b) simple filter ($GR = 1.1$ and $R = 1$), c) Laplace filter ($GR = 1.1$ and $R = 50$), and d) simple filter ($GR = 1.1$ and $R = 50$). Another representation of the transfer functions in the range $R \in (1, 50)$ is provided in figures 2 and 3 of Supplemental Material S1.

Table 3: The first moments of The Laplce filter, simple filter, and new filter for two non-uniform Cartesian grids exhibiting a growth rate of 1.1 in the y-direction, with aspect ratios $R = 1$ and $R = 50$.
$$O_2 = \left((M_x^1)^2 + (M_y^1)^2 + (M_z^1)^2\right)^{1/2}.$$

| Grid | Properties | Laplace filter | simple filter | $F_M$ Filter |
|---|---|---|---|---|
| $R = 1$ and $GR = 1.1$ | $M_x^1$ and $M_z^1$ | 0 | 0 | 0 |
| | $M_y^1$ | 0 | −0.0159 | −0.0159 |
| | $O_2$ | 0 | 0.0159 | 0.0159 |
| $R = 50$ and $GR = 1.1$ | $M_x^1$ and $M_z^1$ | 0 | 0 | 0 |
| | $M_y^1$ | 0 | −0.0458 | −0.0159 |
| | $O_2$ | 0 | 0.0458 | 0.0159 |

Based on the observations made and analyses provided in this section, it is concluded that, among properties 2-7 outlined in section 3.1, The Laplace filter only retains the commutativity property. While its transfer function shape varies across different directions and exhibits a strong dependence on the computational grid configuration. In contrast, the simple filter demonstrates the properties of positivity, stability, and a minimal dispersion error. Nevertheless, similar to the Laplace filter, its transfer function shape considerably changes by direction and is significantly influenced by the grid structure. These findings collectively suggest that neither the Laplace nor simple filter is an ideal candidate for use as an explicit filter in ADM. Both lack the majority of desirable properties enumerated in section 3.1, are highly sensitive to the underlying computational grid and especially the grid aspect ratio, and do not provide a similar behavior across spatial directions.

In light of these limitations, there is a clear need to develop a filter that is applicable to general unstructured grids while is, first and foremost, less sensitive to grid configuration and ensures isotropic properties. More precisely, such a filter should, to the greatest extent possible, satisfy the full range of properties identified in section 3.1.

### 3.3. The new filter definition and analysis

To address the challenge introduced in section 3.2, a novel filter is designed that is independent of the grid aspect ratio, eliminates the need for *ad hoc* boundary treatments, and employs a recursive method to dynamically adjust the filter width. The proposed filter combines a new face-averaging technique with a recursive filtering approach. The formulation of the new filter is as follows:

$$\bar{\phi}_i^n = (1 - b_n)\bar{\phi}_i^{n-1} + b_n F_M(\bar{\phi}_i^{n-1}); n = 1,2,\ldots,N_R, \bar{\phi}_i^0 = \phi_i, \tag{34}$$

$$F_M(\phi_i) = \frac{\sum_{f=1}^{N_{f,i}} (\phi)^f}{N_{f,i}}, \tag{35}$$

where $N_R$ is the number of filter recursions and $b_n$ are the filter relaxation coefficients, constituting $N_R + 1$ filter parameters. To justify why a new face-averaging, $F_M$, is chosen as the base filter instead of $F_S$, Eq. (33), the characteristics of $F_M$ are analyzed and compared with the Laplace and simple filters in this section. In the canonical case of a 3D uniform Cartesian grid with an aspect ratio $R$, where $h_x = h_z = Rh_y$, an alternative form of the filter relation for $F_M$ in Eq. (35) can be expressed by Eq. (21) with the coefficients analytically derived and reported in table 2. Notably, unlike the Laplace and simple filters, the coefficients of the proposed filter are independent of the grid aspect ratio ($R$) and remain identical across all spatial directions.

To further analyze the base filter, in figure 6, the transfer function of the base filter ($F_M$) is examined on four distinct grid configurations: a uniform grid with two aspect ratios of unity and 50, and a non-uniform grid with a growth rate of 1.1 and two aspect ratios of unity and 50 at the target cell. Overall, the real part and magnitude of the transfer function for the new filter exhibit remarkable isotropy across different directions and remain largely unaffected by variations in grid structure. It is also observed that, for both non-uniform grids, the imaginary component of the transfer function in the y-direction remains identical on both non-uniform grids despite differences in the aspect ratio, with a maximum value of approximately 0.025—substantially lower than those of the Laplace and simple filters (9 and 0.08). Furthermore, the first moment of $F_M$ (reported in table 3) remains much smaller than that of the simple filter on a typical channel-flow grid configuration ($GR = 1.1$ and $R = 50$). Nevertheless, the high-wavenumber attenuation property is not fully achieved by this filter, similar to the Laplace and simple filters. To achieve improved filters in terms of satisfying the set of filter properties defined in section 3.1 and more importantly to be able to adjust a prescribed filter width, an optimization problem is formulated next to determine the filter parameters, $N_R$ and $b_n$ ($n = 1, 2, \ldots, N_R$).

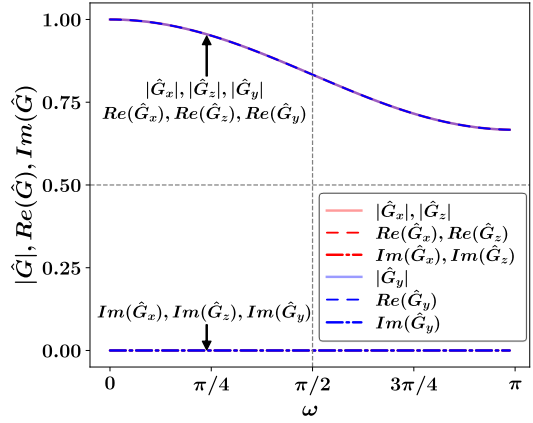

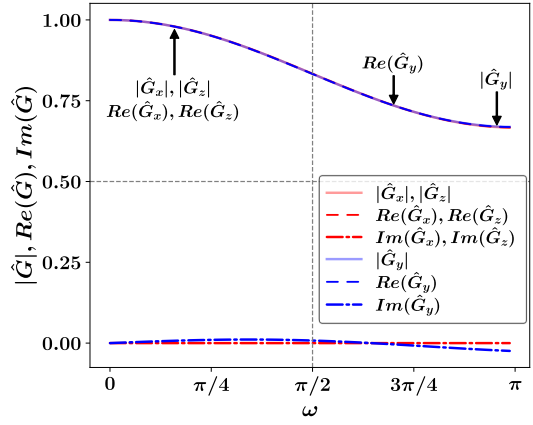

Figure 6: The effects of the grid growth rate ($GR$) and aspect ratio ($R$) on the magnitude, real part, and imaginary part of the base filter ($F_M$) transfer function on different Cartesian grids: a) $GR = 1$ and $R = 1$ or 50, b) $GR = 1.1$ and $R = 1$ or 50.

### 3.4. The new filter optimization procedure

In this section, the parameters of the new filter, including the number of recursions, $N_R$, and relaxation coefficients, $b_n$ ($n = 1, 2, \dots, N_R$), are determined based on a multi-objective optimization process to achieve the desired explicit filter properties introduced in section 3.1. The normalization property, Eq. (24), is automatically satisfied by the definition of the new filter, Eqs. (34) and (35), and thus is not included explicitly as an optimization goal. The normalized objective functions and constraints of the optimization problem are defined in table 4. Note that, here, the wavenumber space is discretized into ten equidistant points in each spatial direction within the range $(0, \pi)$, indexed by $j = 1, \dots, 10$. These points are used to compute the representative values of the transfer function in the wavenumber space.

Table 4: Definition of the normalized objective functions and constraints for the optimization problem. $e_i$ and $\Delta_i$ are the unit vector and filter width in direction $i$.

| Type | Description | Mathematical Formulation | | Count |
|---|---|---|---|---|
| Objective: min $\{O_1\}$ | *High-wavenumber attenuation:* Based on Eq. (25) minimizes the transfer function norm at the Nyquist wavenumber ($\pi$). | $O_1 = \left(\sum_{i=1}^{3}\lvert\hat{G}(\boldsymbol{\omega}=\pi\boldsymbol{e}_i)\rvert^2\right)^{1/2}$ | (36) | 1 |
| Objective: min $\{O_2\}$ | *Commutativity:* Based on Eq. (27) minimizes the norm of the first-order moments to ensure commutation with spatial derivative operators. | $O_2 = \left((M_x^1)^2 + (M_y^1)^2 + (M_z^1)^2\right)^{1/2}$ | (37) | 1 |
| Objective: min $\{O_3\}$ | *Small dispersion error:* Based on Eq. (28) minimizes the imaginary part of the transfer function across 10 representative points per direction. | $O_3 = \left(\sum_{i=1}^{3}\sum_{j=1}^{10}\left[Im\left\{\hat{G}\left(\boldsymbol{\omega}=\frac{j\pi}{10}\boldsymbol{e}_i\right)\right\}\right]^2\right)^{1/2}$ | (38) | 1 |
| Constraint C1: | *Preadjusted filter width:* Constrains the magnitude to ~0.5 at the specific cut-off frequency $\omega_c = h_i\pi/\Delta_i$. | $0.495 \leq \left\lvert\hat{G}\left(\boldsymbol{\omega}=\frac{h_i\pi}{\Delta_i}\boldsymbol{e}_i\right)\right\rvert \leq 0.505$ $i = 1,2,3$ | (39) | 3 |
| Constraint C2: | *Positiveness:* Ensures the real part of the transfer function remains positive at representative points. | $-Re\left\{\hat{G}\left(\boldsymbol{\omega}=\frac{j\pi}{10}\boldsymbol{e}_i\right)\right\} < 0$ $i = 1,2,3; j = 1,2,\ldots,10$ | (40) | 30 |
| Constraint C3: | *Stability:* Ensures the transfer function magnitude does not exceed unity (no amplification) at representative points. | $\left\lvert\hat{G}\left(\boldsymbol{\omega}=\frac{h_j\pi}{\Delta_i}\boldsymbol{e}_i\right)\right\rvert \leq 1$ $i = 1,2,3; j = 1,2,\ldots,10$ | (41) | 30 |

The optimization problem is thus defined by three objective functions $O_1$, $O_2$, and $O_3$, along with 63 nonlinear constraints $C_1$, $C_2$, and $C_3$. The decision variables include $N_R$ and $b_n$ ($n = 1,2,\ldots,N_R$), resulting in a total of $N_R + 1$ decision parameters. Here, $N_R$ is chosen in the range of 3 and 4 (based on computational efficiency considerations), and the range of relaxation $b_n$ is constrained to be $[0,1.5]$ (based on prior testing and practical experience, as it ensures filter stability). Given that $N_R$ is restricted to discrete values, the optimization problem is solved independently for each $N_R$ value, considering $b_n$s as the decision variables only, and the most optimal solution is selected from the optimums for all $N_R$.

To solve this multi-objective optimization problem, MATLAB (www.mathworks.com) and the "gamultiobj" function, using a multi-objective Genetic Algorithm (GA) (Deb 2001), were employed. This approach is particularly well-suited for problems involving multiple competing objective functions, as it efficiently explores the Pareto front to identify optimal trade-offs. We share the optimization code and an example on how to use it for a non-uniform Cartesian grid typical in channel flow problems in Supplementary Material S3. The readers can simply adopt it for their own grid.

To systematically select the optimal solution in a multi-objective optimization framework, we employ a minimum distance-based decision-making function. Given three objective functions $O_i (i = 1,2,3)$ and an ideal solution at $O_i^* = (0,0,0)$, the decision function is formulated to minimize the weighted Euclidean distance from this ideal point. Specifically, we define the decision-making function as follows:

$$D = \left( \sum_{i=1}^{3} \left( w_i (O_i - O_i^*) \right)^2 \right)^{1/2}, \tag{42}$$

where $w_i$ represents the assigned weight for each objective, ensuring a balanced trade-off among conflicting criteria. In this study, we assume equal importance of objectives, i.e., $w_i = (1,1,1)$, while other weights can be simply adopted by the users.

### 3.5. The new filter optimization results

The ALDME model necessitates a filter width of two times the grid size, serving as both the explicit convolution $\overline{(.)}$ and test $\widetilde{(.)}$ filters. Therefore, here we design an explicit filter of width $\Delta_i/h_i = 2$, necessary for the implementation of the AD-LES model described in section 2. Nevertheless, the design algorithm and provided code described in section 3 can also be employed for other filter widths. It is important to note that the transfer function depends not only on the filter parameters but also on the characteristics of the computational grid. In wall-bounded flows, the grid is typically refined near the wall and coarsens at a specified growth rate as the distance from the wall increases. The optimization is performed on a non-uniform Cartesian grid with a growth rate of 1.1 in the y-direction and an aspect ratio of 50 near the walls, a typical choice in LES applications (Gullbrand 2004, Sagaut 2005, Choi & Moin 2012, Singh & You 2013, Schumann *et al.* 2020, van Druenen & Blocken 2024). The new filter coefficients are independent of the grid aspect ratio (see section 3.3), therefore, this grid configuration ensures that the optimized filter performs effectively across the range of scales encountered in practical simulations. The results of the optimization process are summarized in table 5. Additionally, the magnitude, real, and imaginary parts of the transfer functions for filters with $N_R$ values of 3 and 4 are illustrated in figure 7. As evident from the results, the transfer functions for both $N_R$ values exhibit nearly identical characteristics. Therefore, when considering the computational cost, the most optimal filter is achieved by choosing $N_R = 3$ and $b_n = [1.2117, 1.2344, 1.2189]$. It should be noted that the result of a GA optimization is prone to a level of variation and uncertainty, depending on the initialization, chosen algorithm parameters' values, randomness of different steps, and the complexity of the problem. To ensure the validity of the optimization results, we conducted an uncertainty analysis by repeating the optimization process 100 independent times. The filter coefficients ($b_n$) showed a minor variation with a standard deviation of approximately 0.02 (representing a variation of ~1.7% relative to the mean), and the objective function values ($O_1, O_2, O_3$) exhibited standard deviations in the order of $10^{-5}$ to $10^{-6}$.

Table 5: The optimal filter parameters of the new filter for a typical channel-flow grid (a non-uniform Cartesian grid with $GR = 1.1$ and $R = 50$).

| $N_R$ | $b_1$ | $b_2$ | $b_3$ | $b_4$ | $O_1$ | $O_2$ | $O_3$ |
|---|---|---|---|---|---|---|---|
| 3 | 1.2117 | 1.2344 | 1.2189 | - | 0.3607 | 0.0583 | 0.0687 |
| 4 | 0.9581 | 0.9719 | 0.9225 | 0.9182 | 0.3830 | 0.0600 | 0.0662 |

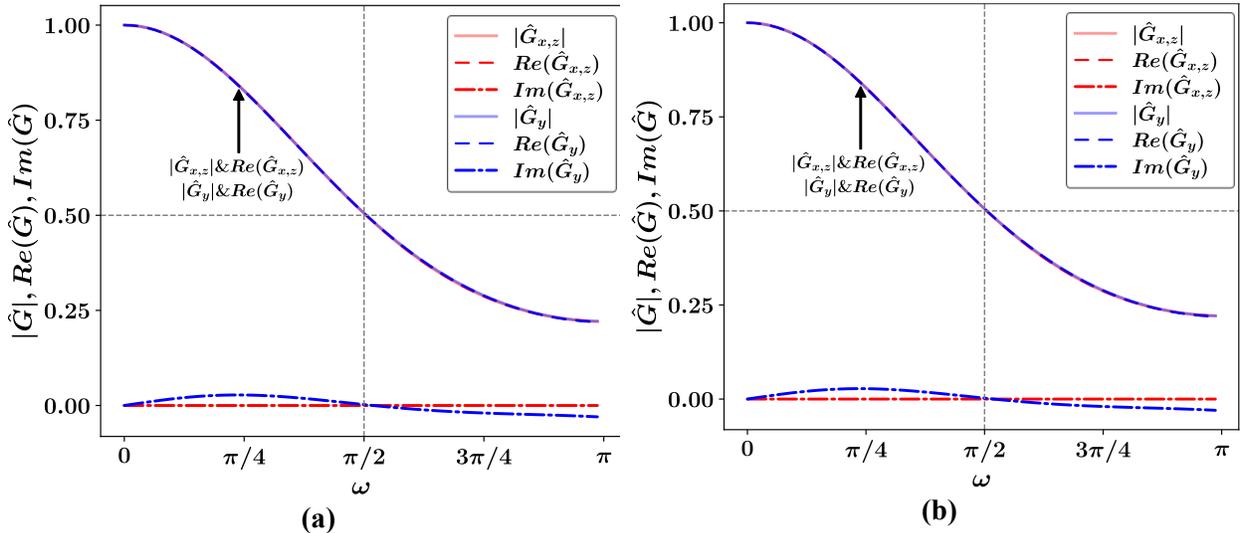

Figure 7: The magnitude, real part, and imaginary part of the transfer functions of the optimized new filter corresponding to the filter parameters given in table 5: a) $N_R = 3$ and b) $N_R = 4$.

It is worth noting that the divergence-preserving property of an explicit filter mathematically relies on its commutativity ($\overline{\partial_i u_i} \approx \partial_i \bar{\tilde{u}}_i$), which was accounted in this work. However, another condition on the sum of fluxes at the faces of computational cells (Agdestein & Sanderse 2025) is also necessary to ensure the divergence-preserving property ($\partial_i \bar{\tilde{u}}_i = 0$), which was not considered here.

### *3.6. The new filter optimization for tetrahedral grids*

Since the new filter is applicable to general unstructured grids, we investigate the filter parameters and performance for general tetrahedral unstructured grids in this section. To assess the performance of the new filter on unstructured grids, its transfer function is evaluated over two tetrahedral grid configurations. The first grid (shown in figure 8a) exhibits an aspect ratio of 1.23, a mean volume ratio of 1.25, and a skewness equiangle of 0.386, while the second grid (shown in figure 8b) which resembles a boundary layer grid has an aspect ratio of 5.68, a mean volume ratio of 1.34, and a skewness equiangle of 0.934. For this analysis, the optimization (for $N_R = 3$) yields the filter relaxation parameters of $b_n = [1.0786, 1.0765, 1.0829]$ for the first grid and $b_n = [0.9968, 1.0130, 1.0199]$ for the second one.

The MATLAB optimization code, the inputs and settings for these two unstructured tetrahedral grids, and a succinct user guide are available in Supplementary Material S4. This resource includes instructions for utilizing the code as well as guidance on executing the scripts. Figure 9a and b present the transfer functions of the filters with the optimized coefficients. Notably, despite the relatively low quality of the computational grid, the filter maintains close to isotropic, except for the dispersion error, with a reasonably transfer function properties across each spatial direction.

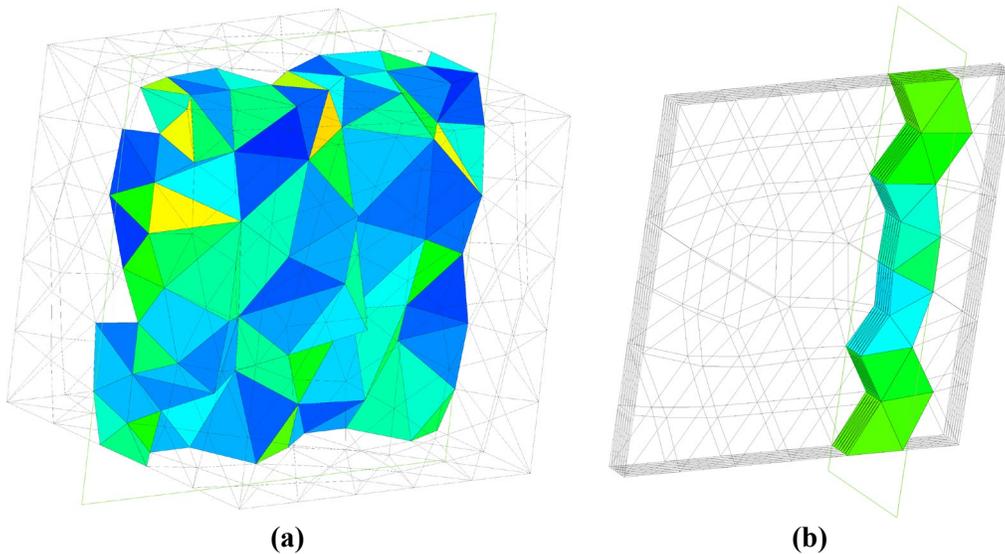

(a)          (b)
Figure 8: Visualization of the two unstructured tetrahedral grids: a) The first grid with an aspect ratio close to unity, and b) the second grid with a large aspect ratio.

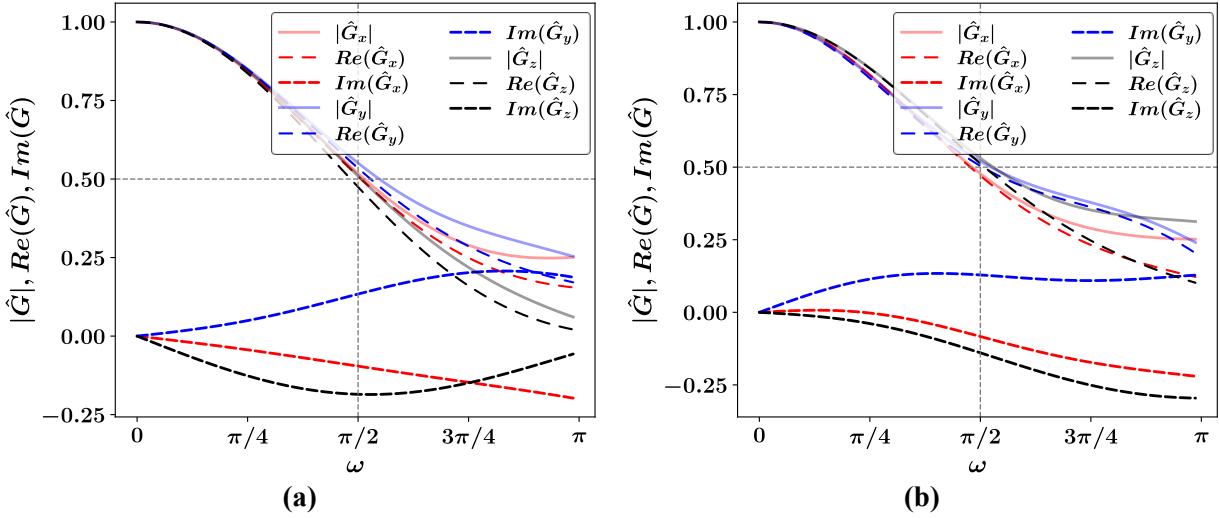

(a) (b)

Figure 9: The magnitude, real part, and imaginary part of the optimal new filter transfer function over: a) the first tetrahedral grid, and b) the second tetrahedral grid.

## 4. *The a posteriori* analysis of the new filter

This section evaluates the performance of the proposed filter for AD-LES using a general unstructured grid code, i.e., OpenFOAM.

### *4.1. New filter performance on highly stretched boundary-layer grids*

In the chosen AD-LES closure, i.e., ALDME (see section 2), both the deconvolution filter and the test filter have a width of $2\widetilde{\Delta} = 2h$, where $\widetilde{\Delta} = h = \sqrt[3]{V}$ and $V$ is the local cell volume. Consequently, the filter parameters obtained from the optimization process in section 3.5—specifically, $N_R = 3$ and $b_n = [1.2182, 1.2088, 1.2381]$ —are adopted for the current simulations.

For each of the channel flow benchmarks at $Re_\tau = 395$ and $Re_\tau = 590$, three AD-LES simulations are carried out, each using one of the FVM explicit filters investigated in this work for general unstructured grids, i.e., The Laplace filter, simple filter, and newly developed filter (newFilter). All simulations using the Laplace filter diverged. Just before divergence, strong unphysical oscillations of the velocity were observed near the wall boundaries, similar to those occurred in our *a priori* analysis (see figure 3). The reason behind this issue of the Laplace filter was analyzed in detail in section 3.2. Figure 10 and figure 11 present the flow statistics, including the mean velocity and Reynolds stress profiles, predicted using the new filter and widely-used simple filter against the reference DNS data. The results using the new filter demonstrates noticeable improvements compared to the one with the simple filter, specifically for the mean velocity profile. The amount of log-layer mismatch is alleviated by the use of the new filter. It is worth noting that some previous studies, e.g., Schlatter et al. (2004), often reported limited improvements with conventional AD-LES closures compared to pure DEV-LES. Through theoretical analyses in our recent work (Amani et al. 2024), we demonstrated that existing conventional penalty-term AD-LES and even standard mixed AD-LES suffer from inherent mathematical inconsistencies. Then, we proposed novel mixed AD-LES closures and proved their advantages in several *a posteriori* tests. The consistent mixed approaches can combine the best of both worlds: the high-fidelity structural reconstruction of AD and the stability of EV, yielding results that are superior to conventional methods. From those consistent mixed AD-LES models, the best one has been chosen in the present work to investigate the effect of explicit filter design. The result of a widely-used pure EV-LES, namely WALE, is also reported in figure 10a. The comparison provided in this figure highlights that the consistent mixed AD-LES model improves the flow predictions, more noticeably when a proper explicit filter such as the one designed in the present work is adopted. In other word, the explicit filter design is the key element of AD-LES to achieve high fidelity results.

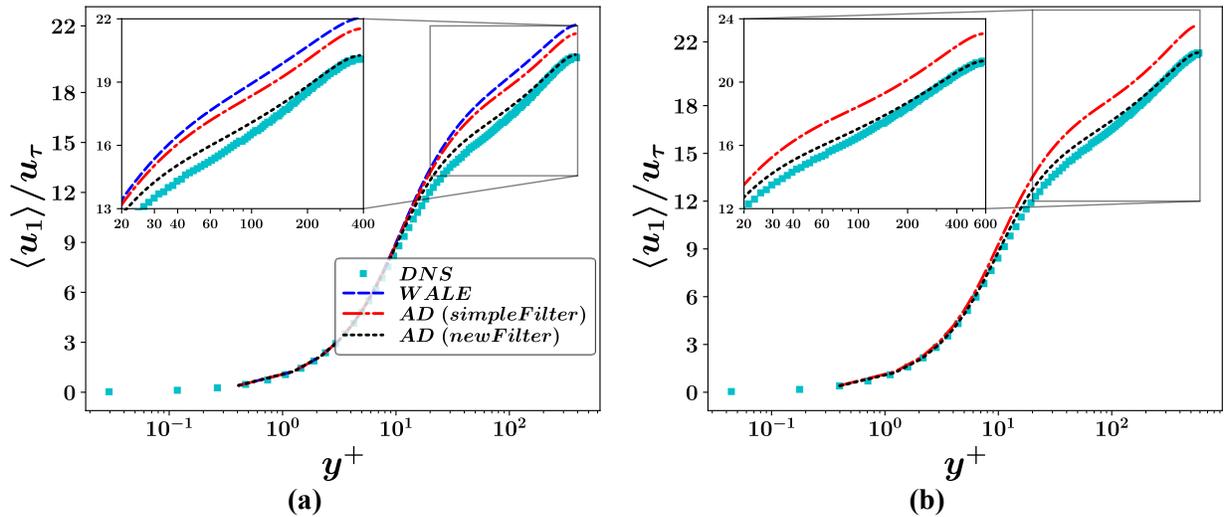

Figure 10: The (dimensionless) mean streamwise velocity profiles: A comparison of the AD-LES results using the new and simple filter against the DNS data (Moser et al. 1999): (a) $Re_\tau = 395$, and (b) $Re_\tau = 590$. The AD-LES using the Laplace filter diverged. Part (a) includes the results of the pure eddy-viscosity WALE as a reference LES.

As shown in figure 11, using the new filter, improvements are also observed in the prediction of the streamwise Reynolds stress, which is the dominant normal Reynolds stress component, and the shear stress, specifically at higher Reynolds numbers ($Re_\tau = 590$). However, a level of under-prediction in the wall-normal Reynolds stress component, $\langle u'_2 u'_2 \rangle$, is evidenced with the new filter at the higher Reynolds number (figure 11d). Notably, as shown by Fröhlich and Rodi (2002), the high over-prediction of the peak $\langle u'_1 u'_1 \rangle$ value (figure 11a and b) in LES results of channel flow may be connected to the grid size in the spanwise direction, and the reduction of this factor can effectively mitigate this error.

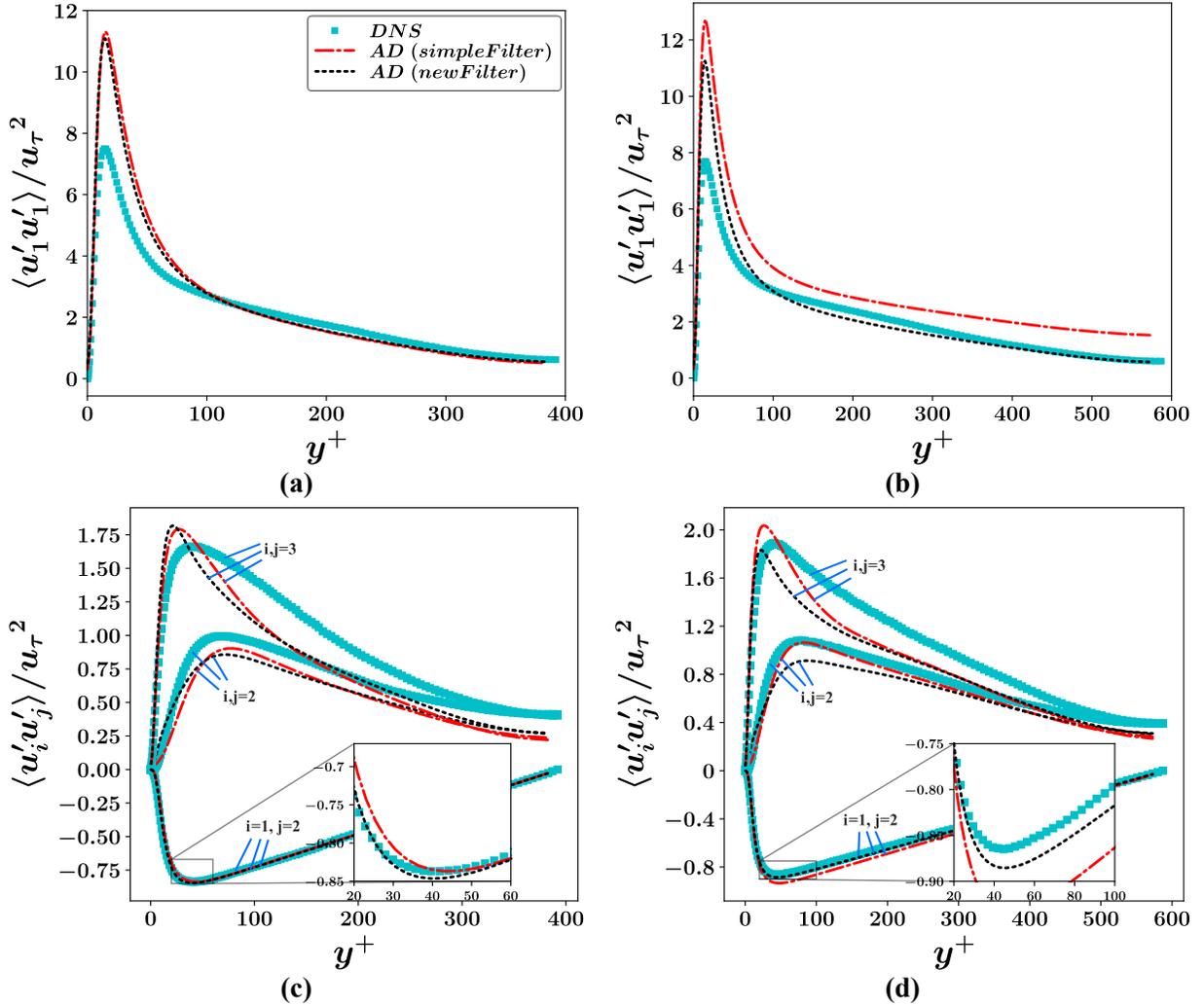

Figure 11: The (dimensionless) Reynolds stresses: A comparison of AD-LES results using the new and simple filters against the DNS data (Moser et al. 1999): Left column: $Re_\tau = 395$, and right column: $Re_\tau = 590$. The AD-LES using the Laplace filter diverged.

To explore the origin of these enhancements using the new filter in greater details, the components of the SFS stresses, i.e., $b_{ij}$ and $a_{ij}$ in Eq. (6), are plotted in figure 12. According to these figures, using the new filter leads to a significant increase in the values of these two stresses. For a more quantitative analysis, table 6 reports the contributions of various terms to the turbulent kinetic energy. The data in this table further support the earlier observation of increased SFS stresses. On the other hand, applying the new filter results in approximately a 30% reduction in the resolved kinetic energy. This is justified by the comparison of the shape of the transfer functions of the new filter (figure 7a) and simple filter (figure 5d). There is an insufficient level of attenuation near the Nyquist in the x- and z-directions using the simple filter. On the other hand, the new filter operates isotropically and offers a reasonable attenuation in all three directions. As a result, the level of directly resolved kinetic energy declines and the resolved part using deconvolution increases using the new filter. Therefore, it is inferred that filtering out the frequencies near the Nyquist (sufficient attenuation near the Nyquist) is an important feature of an ADM explicit filter, which is effectively achieved in all directions by the new filter and brings about more accurate predictions. It is worth mentioning that almost all of this reduction in the directly resolved kinetic energy is recovered by the deconvolution (deconvolved kinetic energy in table 6) and the fraction of the modeled kinetic energy still remains below 10% for the new filter in both cases, reflecting well-resolved LES based on the well-known Pope's criterion (Pope 2000). To further show that the incorporation of the new filter in our study meets the criteria for a high-quality LES, the LES quality criterion based on the index of resolution (Celik *et al.* 2005), namely the ratio of the eddy viscosity to the molecular viscosity, is examined in figure 13. It can be seen that this ratio is well below the critical value of 20, for a well-resolved LES, in both benchmark cases using the new filter.

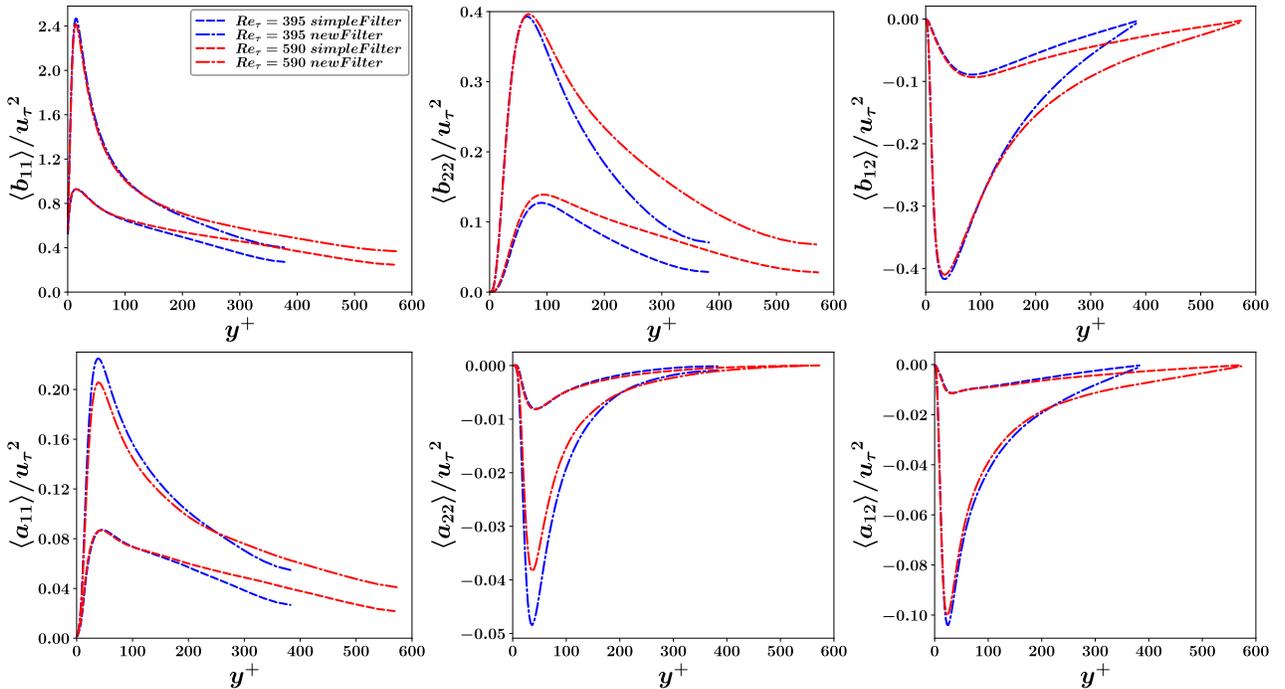

Figure 12: A comparison of the deconvolved, $b_{ij}$, (top row) and modeled, $a_{ij}$, (bottom row) SFS stress components using the new and simple filters.

Table 6: A comparison of the kinetic energy budget (in percent) of the directly resolved, deconvolved, and modeled stress using the new and simple filters.

| Benchmark | Filter | $\frac{1}{2}\langle \bar{\tilde{u}}_i'\bar{\tilde{u}}_i' \rangle$ (Directly resolved) | $\frac{1}{2}\langle b_{ii} \rangle$ (Deconvolved) | $k_{sgs}$ (Modeled) |
|---|---|---|---|---|
| $Re_\tau = 395$ | simpleFilter | 81.83 | 15.2 | 2.97 |
|  | newFilter | 49.17 | 43.7 | 7.12 |
| $Re_\tau = 590$ | simpleFilter | 85.62 | 12.02 | 2.36 |
|  | newFilter | 56.22 | 38.15 | 5.63 |

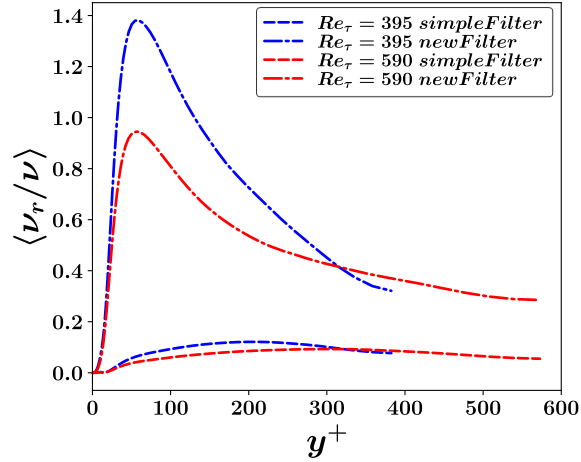

Figure 13: A comparison of the eddy viscosity ratio profiles using the new and simple filters.

### 4.2. New filter performance on unstructured grids

For *a posteriori* tests of unstructured grids in the TGV benchmark, as a simplification, the optimal parameters obtained in section 3.6—specifically, $N_R = 3$ and $b_n = [1.0786, 1.0765, 1.0829]$— are used in this section. The computational cell size of the reference DNS solution (DeBonis 2013) is denoted by $\Delta_{DNS}$. For the current LES, unstructured grids (figure 2) composed of prismatic cells (with hexagonal bases in the x-y planes and prism axes in the z-direction) and different aspect ratios are generated to study this benchmark. The cell base size is indicated by $\Delta_{XY}$ and the height size by $\Delta_Z$. Note that the prism cell topology is used in place of a tetrahedral one, since the former is more suited for LES, especially using stretched grid cells, to avoid too large dissipation errors.

Figure 14 shows iso-surfaces of the Q-criterion predicted by the present AD-LES on an unstructured grid. Figure 15 presents a comparison of the results of AD-LES on unstructured grids with different cell aspect ratios, using both the simple and new filters, against DNS data. For the grid with a low aspect ratio ($\Delta_Z/\Delta_{DNS}= 4, \Delta_{XY}/\Delta_{DNS}= 2$) and the highest resolution close to the DNS, the predicted kinetic energy evolutions in time are in close agreement with the DNS. This highlights the consistency of the solver and solution, using both filters, in the limit where the resolution approaches to the one of DNS. It should be emphasized that while the LES grid resolution is very close to the DNS in this case, the latter possesses higher spatial and temporal discretization schemes. As the grid resolution decreases, by increasing the cell aspect ratio (stretch), the deviations of the LES results from the DNS data grow for both filters. However, at the highest value of the stretch ($\Delta_Z/\Delta_{DNS}= 16, \Delta_{XY}/\Delta_{DNS}= 2$), the result using the new filter demonstrates a much closer agreement with the DNS in the second half of the simulation period, while in the first half, the results of both filters are very close together. This can be justified by noting that the overall size of the vortical structures, in this decaying benchmark, reduces by time (see figure 14). Considering the smaller vortical structures in the second half of the simulation, the effects of the filter cut-off length and properties are deemed much more important in the second period.

Finally, it is worth noting that in this section, as a simplification, the optimal parameters obtained for tetrahedral cells are used for prism cells. This simplification is made owing to the low sensitivity of the filter coefficients observed in section 3.6. Further examination of this simplification and obtaining the filter coefficients for different cell topologies would be a valuable topic for future research.

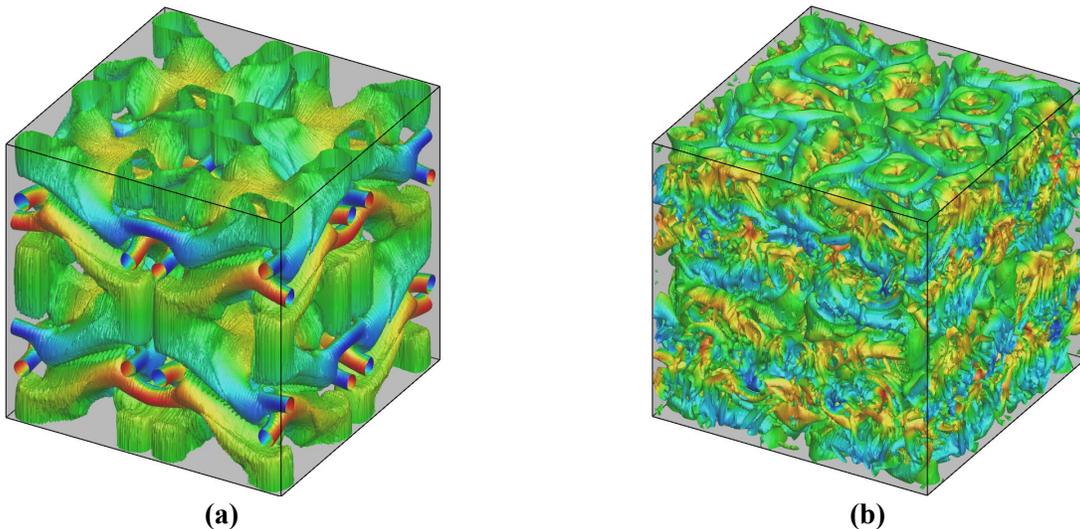

(a)  (b)
Figure 14: The TGV benchmark: Q-criterion iso-surfaces predicted by the present AD-LES on an unstructured grid ($\Delta_Z/\Delta_{DNS}= 4, \Delta_{XY}/\Delta_{DNS}= 2$) at time $tV_0/L = 5$ (a) and 15 (b).

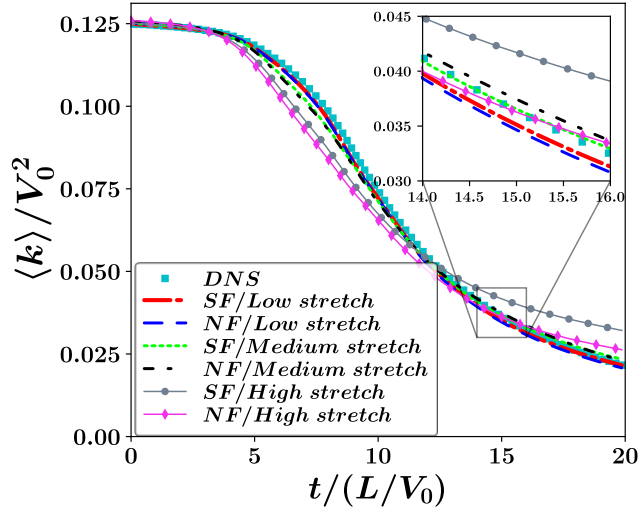

Figure 15: The TGV benchmark: The (dimensionless) volume-averaged total kinetic energy versus dimensionless time. The comparison of the DNS solution (DeBonis 2013) with the present AD-LES results using different filters—Simple Filter (SF) and New Filter (NF)—and grid aspect ratios: low stretch ($\Delta_Z/\Delta_{DNS} = 4, \Delta_{XY}/\Delta_{DNS} = 2$), medium stretch ($\Delta_Z/\Delta_{DNS} = 8, \Delta_{XY}/\Delta_{DNS} = 2$), high stretch ($\Delta_Z/\Delta_{DNS} = 16, \Delta_{XY}/\Delta_{DNS} = 2$). A zoomed-in view around $tV_0/L = 15$ is shown in the inset.

## 5. Conclusion

This study aimed to develop and evaluate a novel optimized recursive explicit filter for AD-LES applicable to unstructured grids. To this end, detailed analyses were first conducted on two commonly used filters for general unstructured grids—namely, a differential filter (laplaceFilter) and a face-averaging filter (simpleFilter)—across various canonical test cases. It was demonstrated that both filters exhibit strong directional anisotropy in their transfer function shape and acute performance sensitivity to grid configuration, especially to the cell aspect ratio. Both filters showed inadequate attenuation at high wavenumbers in all or some spatial directions on typical channel-flow grids. In addition, the Laplace filter violates the filter stability and positivity criteria on grids with moderate to high aspect ratios, eventually leading to solution divergence. On the other hand, for the simple filter, the commutativity error grows with the increase in the aspect ratio. To address these shortcomings, we proposed a new filter, first and foremost, to remove the strong sensitivity of the filter coefficients and transfer function to the grid aspect ratio and deliver an isotropic transfer function. For this purpose, a recursive filtering approach and a face-averaging technique were combined. Then, through a constrained multi-objective optimization, the filter parameters were determined to precisely adjust a prespecified filter width and maintain the filter stability and positivity, while maximizing the high-wavenumber attenuation and minimizing the dispersion and commutativity errors. Comparative *a priori* analyses in several test cases, including uniform

Cartesian grids of different aspect ratios, non-uniform Cartesian grids of various aspect ratios, and unstructured tetrahedral grids with small and large aspect ratios, revealed substantial improvements in the properties of the new filter over those of the conventional filters. Furthermore, the proposed filter obviated the need for additional boundary treatments and offered much more parallel computational efficiency compared to differential filters. Finally, we conducted *a posteriori* analyses through AD-LES of two turbulent channel flow benchmarks. While all simulations satisfied the criteria of a well-resolved LES, the mean velocity prediction using the new filter showed noticeable improvements over those using the other filters, and there were also slight enhancements in the prediction of the normal streamwise and shear Reynolds stress components. This progress was primarily linked to the sufficient and isotropic high-wavenumber attenuating property of the new filter for an AD-LES. Finally, to show the application of the new filter on more general unstructured grids, *a posteriori* tests on the 3D TGV benchmark using prism cells were carried out successfully. Future directions for extending the present study include further enhancements to the filter properties, e.g., by accounting for the divergence-preserving property, and the test of the present filter performance on different unstructured-grid cell topologies, which necessitates the use of the proposed algorithm to obtain the optimized filter coefficients and the design of unstructured-grid *a priori* or *a posteriori* benchmarks.

**Supplementary materials**

The supplemental materials include additional graphs of filter properties (Supplementary material S1), the computer programs and user guides for the calculation of the filter coefficients on arbitrary grids (Supplementary material S2), the optimization code for a non-uniform Cartesian grid (Supplementary material S3), and the optimization code for unstructured tetrahedral grids (Supplementary material S4).

**Data availability**

The data that support the findings of this study are available from the corresponding author upon reasonable request.

**Declaration of Interests:** The authors report no conflict of interest

**Acknowledgments**

This work was partially supported by TÜBITAK [grant number 221M421].

**Appendix A: The 2nd-order Padé filter coefficients**

The general one-dimensional form of the second-order Padé filter (Lele 1992) on a uniform grid is expressed as follows:

$$\alpha \bar{\phi}_{i-1} + \bar{\phi}_i + \alpha \bar{\phi}_{i+1} = a\phi_i + \frac{b}{2}(\phi_{i-1} + \phi_{i+1}), \tag{43}$$

where $\alpha = -\frac{1}{2}\cos(\omega_c)$ and $a = b = \frac{1}{2} + \alpha$. For a non-dimensional cut-off wavenumber of $\omega_c = \pi/2$, Eq. (43) is simplified to:

$$\bar{\phi}_i = \frac{1}{2}\phi_i + \frac{1}{4}(\phi_{i-1} + \phi_{i+1}). \tag{44}$$

When this one-dimensional filter is applied independently along each of the three principal directions on a uniform Cartesian grid, the resulting coefficients are identical to those presented in table 2.

**Appendix B: Stability criterion of the Laplace filter on uniform Cartesian grids of different aspect ratios**

To assess the stability of the Laplace filter on a uniform Cartesian grid, its transfer function is analytically derived in the $x$, $y$, and $z$ directions, and its stability is examined based on Eq. (30). Specifically, the transfer function given by Eq. (22) is expanded for each spatial direction. For instance, the transfer function in the x-direction becomes:

$$\hat{G}_x = \hat{G}\left(\mathbf{k} = \left(\frac{\omega_x}{h_x}, 0, 0\right)\right) = \sum_{j=1}^{N_c} a_{i,j} e^{-i\left(\frac{\omega_x}{h_x}, 0, 0\right) \cdot (x_i - x_j)} = \tag{45}$$
$$= a_{i,j,k} + a_{i+1,j,k} e^{-i\frac{\omega_x}{h_x} h_x} + a_{i-1,j,k} e^{i\frac{\omega_x}{h_x} h_x} + a_{i,j+1,k} + a_{i,j-1,k} + a_{i,j,k+1} + a_{i,j,k-1}.$$

Substituting the filter coefficients, $a_{i,j,k}$, on a uniform grid with an aspect ratio of $R$ in the y-direction from table 2, Eq. (45) is reduced to

$$\hat{G}_x = 1 + \frac{2}{cR^{\frac{2}{3}}}(\cos\omega_x - 1). \tag{46}$$

Applying the stability criterion, Eq. (30), yields:

$$-1 \leq \left[1 + \frac{2}{cR^{\frac{2}{3}}}(\cos\omega_x - 1)\right] \leq 1. \tag{47}$$

Since $c$ and $R$ are always positive, the second inequality is always held for all $\omega_x$ while the first one is true if

$$cR^{\frac{2}{3}} \geq 2. \tag{48}$$

If $c > 2$, the above condition is always met since $R > 1$ is assumed. By a similar argument, it can be shown that for $\hat{G}_z$, the stability condition is the same as Eq. (48).

Applying a similar approach for the transfer function in the y-direction, $\hat{G}_y$, yields the following constraint:

$$\frac{c}{R^{\frac{4}{3}}} \geq 2 \text{ or } R \leq \left(\frac{c}{2}\right)^{\frac{3}{4}}. \tag{49}$$

Based on Eq. (49), the Laplace filter with a coefficient of $c$ is stable only when the aspect ratio of the grid is less than or equal $(c/2)^{\frac{3}{4}}$.

## Appendix C: The calculation of the filter coefficients on arbitrary grids

For the calculation of the filter coefficients on arbitrary grids—specifically tetrahedral and hexahedral grids— computational grids data were first imported into MATLAB. A symbolic field was then constructed to represent a field values at each computational cell, upon which the filtering operation was applied symbolically to the target cell. By leveraging the linear dependence of the filtered value on neighboring cell values, the filter weighting coefficients ($a_{ij}$) were systematically derived through a single symbolic differentiation step. These coefficients were subsequently utilized to compute the filter transfer function as defined by Eq. (22). The MATLAB scripts, along with a concise user guide for utilizing this code, are provided in Supplementary Material S2. This inclusion is intended to facilitate ease of use for readers, enabling them to effectively adopt the scripts in their research endeavors. To ensure the validity of the MATLAB computations, the derived filter coefficients for the canonical case of a uniform grid with an aspect ratio of $R$ were validated against the analytically derived values reported in table 2.